\RequirePackage{fix-cm}
\documentclass[preprint]{elsarticle}
\usepackage{lineno}
\modulolinenumbers[5]

\usepackage{graphicx}
\usepackage{amsmath}
\usepackage{amsfonts}
\usepackage{xcolor,colortbl}
\usepackage{subfiles}
\usepackage{url}
\usepackage{subfig}
\usepackage{tabularx}
\usepackage{multirow}
\usepackage{xcolor}
\usepackage[final]{changes}
\usepackage{hhline}

\definechangesauthor[name={Second Version}, color=blue]{V2}
\colorlet{Changes@Color}{red}

\DeclareMathOperator*{\argmin}{arg\,min}
\DeclareMathAlphabet\mathbfcal{OMS}{cmsy}{b}{n}
\newcommand{\T}{\mathcal{T}}

\definecolor{Gray}{gray}{0.92}
\newcolumntype{g}{>{\columncolor{Gray}}l}

\begin{document}
\begin{frontmatter}

\title{Precise Estimation of Renal Vascular Dominant Regions Using Spatially Aware Fully Convolutional Networks, Tensor-Cut and Voronoi Diagrams}
%\thanks{Grants or other notes
%about the article that should go on the front page should be
%placed here. General acknowledgments should be placed at the end of the article.}

%\subtitle{Do you have a subtitle?\\ If so, write it here}
%\titlerunning{Renal Vascular Dominant Region Estimation}  % if too long for running head

%\author{Chenglong Wang \and
%		Holger R.Roth \and
%        Takayuki Kitasaka \and
%        Masahiro Oda \and
%        Hayashi Yuichiro \and
%        Yasushi Yoshino  \and 
%        Tokunori Yamamoto \and 
%        Kensaku Mori 
%}

%\institute{ Chenglong Wang\at
%              Graduate School of Information Science, Nagoya University, Naogya, Japan
%              \email{cwang@mori.m.is.nagoya-u.ac.jp}
%           \and
%           Takayuki Kitasaka \at
%           School of Information Science, Aichi Institute of Technology \and
%           Yasushi Yoshino, Tokunori Yamamoto \at 
%           Nagoya University Graduate School of Medicine, Nagoya, Japan \and
%            Holger R.Roth, Masahiro Oda, Hayashi Yuichiro, Kensaku Mori \at
%           Graduate School of Informatics, Nagoya University, Nagoya, Japan 
%}

\author[a]{Chenglong Wang\corref{correspondingauthor}}
\author[b]{Holger R. Roth}
\author[c]{Takayuki Kitasaka}
\author[b]{Masahiro Oda}
\author[b]{Yuichiro Hayashi}
\author[d]{Yasushi Yoshino}
\author[d]{Tokunori Yamamoto}
\author[d]{Naoto Sassa}
\author[d]{Momokazu Goto}
\author[b]{Kensaku Mori\corref{correspondingauthor}}

\cortext[correspondingauthor]{{cwang@mori.m.is.nagoya-u.ac.jp} or {kensaku@is.nagoya-u.ac.jp}}

\address[a]{Graduate School of Information Science, Nagoya University, Nagoya, Japan}
\address[b]{Graduate School of Informatics, Nagoya University, Nagoya, Japan}
\address[c]{School of Information Science, Aichi Institute of Technology}
\address[d]{Nagoya University Graduate School of Medicine, Nagoya, Japan}

%\date{Received: date / Accepted: date}

%\maketitle

\begin{abstract}
This paper presents a new approach for precisely estimating the renal vascular dominant region using a Voronoi diagram. To provide computer-assisted diagnostics for the pre-surgical simulation of partial nephrectomy surgery, we must obtain information on the renal arteries and the renal vascular dominant regions. We propose a fully automatic segmentation method that combines a neural network and tensor-based graph-cut methods to precisely extract the kidney and renal arteries. First, we use a convolutional neural network to localize the kidney regions and extract tiny renal arteries with a tensor-based graph-cut method. Then we generate a Voronoi diagram to estimate the renal vascular dominant regions based on the segmented kidney and renal arteries. {The accuracy of kidney segmentation in 27 cases with 8-fold cross validation reached a Dice score of 95\%. The accuracy of renal artery segmentation in 8 cases obtained a centerline overlap ratio of 80\%}. Each partition region corresponds to a renal vascular dominant region. The final dominant-region estimation accuracy achieved a Dice coefficient of 80\%. A clinical application showed the potential of our proposed estimation approach in a real clinical surgical environment. \added{Further validation using large-scale database is our future work.}
\end{abstract}


\end{frontmatter}

% use \includeonly{} for debug
\section{Introduction}\label{intro}
Partial nephrectomy (PN), which has recently become one of the most common treatments for kidney cancer, can maintain a high residual renal function during surgery \cite{shao2011laparoscopic,shao2012precise,yoshino2015}. A critical problem during PN is that blood vessel clamping directly influences the quality of the surgery. However, PN surgery remains unstandardized. Due to the trade-off between residual renal function and surgical difficulty, it is difficult to design a criteria for PN surgery. In this work, we provide a better and more accurate computer-assisted diagnosis for PN by estimating the dominant region of each renal artery that facilitates identifying the blood vessels which feed the tumor. Physicians can easily make a surgical plan using such diagnosis information to determine the blood vessel clamping.

The feasibility of computer-aided diagnosis (CAD) for PN has been proven \cite{ukimura2012three,komai2014novel,isotani2015feasibility}. Ukimura et. al \cite{ukimura2012three} performed PN on four patients who underwent 3D reconstruction for surgical navigation. The kidney surface was extracted by thresholding, the renal arteries were segmented by a simple region-growing method, and the tumor was manually segmented. Komai et al. \cite{komai2014novel} and Isotani et al. \cite{isotani2015feasibility} used commercial software called ``Vincent'' to perform the computer analysis for PN clamping. The kidney was extracted by a semi-automatic region-growing method, and the renal arteries were segmented by applying facial detection technology using multi-phase information. Then the vascular dominant regions were estimated by applying a Voronoi diagram. All of the above research focused on the clinical study of CAD's feasibility and its accuracy for PN surgery rather than engineering studies.

Precise estimation of the renal vascular dominant regions will contribute to more precise surgery, especially for PN surgeries. To precisely estimate the renal vascular dominant regions, precise kidney and renal artery segmentation is essential. 
In this work, we used a deep learning technique to extract the kidney and a tensor-based graph-cut approach to precisely segment the renal arteries. Finally, we used the automatically extracted kidney and renal arteries to estimate the vascular dominant regions using a Voronoi diagram.  

Many organ segmentation methods have already been presented in the literature over the years. Statistical model methods are commonly used in abdominal organ segmentation problems. Lin et al. presented an elliptic candidate region to localize kidneys \cite{lin2006computer}. Zhou et al. constructed a prior shape model from a statistical atlas map for a liver segmentation problem \cite{zhou2006constructing}. Heimann et al. proposed deformable statistical shape models (SSMs) using atlas information \cite{heimann2009statistical}. Hybrid methods have also been proposed. Skalski et al. used a level-set method with ellipsoidal shape constraints for kidney segmentation \cite{skalski2017kidney}. Okada et al. used a combination of SSMs and an intensity model for a multi-organ segmentation problem \cite{okada2015abdominal}. Graphical models have also been widely used for organ segmentation. In our previous work \cite{wang2016precise}, we used a graph-cut method to semi-automatically segment the kidney. Freiman et al. combined shape and graphical models to automatically extract the kidney region \cite{freiman2010non}.

Machine learning techniques have also been applied to organ segmentation problems. Cuingnet et al. proposed a coarse-to-fine kidney segmentation method using random forests \cite{cuingnet2012automatic}. Support vector machines (SVMs) have also benn used for organ segmentation \cite{akbari2012automatic,liu2007learning}. Recently, researchers have adapted deep learning techniques to organ segmentation. Many state-of-the-art results have been achieved in the lungs \cite{harrison2017progressive}, the pancreas \cite{roth2018spatial}, the liver \cite{christ2016automatic}, the head and neck \cite{zhu2018anatomynet}, the mammogram mass \cite{zhu2018adversarial}, and the kidneys \cite{thong2016convolutional,zheng2017deep}.

The segmentation of renal arteries is another critical step in this work. Unlike other tissues like the liver, renal arteries have lower contrast that complicates the extraction of tiny blood vessels. Even though Hessian-based vesselness enhancement filters have been widely used in tubular structure segmentation \cite{sato,frangi}, they are unable to extract tiny blood vessels, especially of low contrast. Friman et al. presented novel template model tracking with a multiple hypothesis procedure \cite{friman}. Multiple hypothesis tracking schemes solved the early termination problem, but specifying a global terminal threshold is difficult. It thus leads to serious over-segmentation, especially for tiny blood vessels of low contrast. Recently, a deep learning technique extracted retinal blood vessels \cite{fu2016retinal,liskowski2016segmenting,fu2016deepvessel}, showing its high potential for blood vessel segmentation. Since the deep learning technique is a data-demanding approach, it also requires large-scale annotated data for supervised learning. However, creating pixel-wise ground-truth labels for big-data is very labor-intensive, especially for 3D medical data.

The main contributions of this paper include the presentation of a fully automated and precise renal vascular dominant estimation approach using a deep learning method for kidney segmentation and a tensor-based graph-cut method for renal artery segmentation. \added{As a preliminary study on CAD system for PN surgery, this work shows the potential of using medical image-processing techniques in improving precise PN surgeries. }
\begin{figure}[t]
  \includegraphics[width=0.9\linewidth]{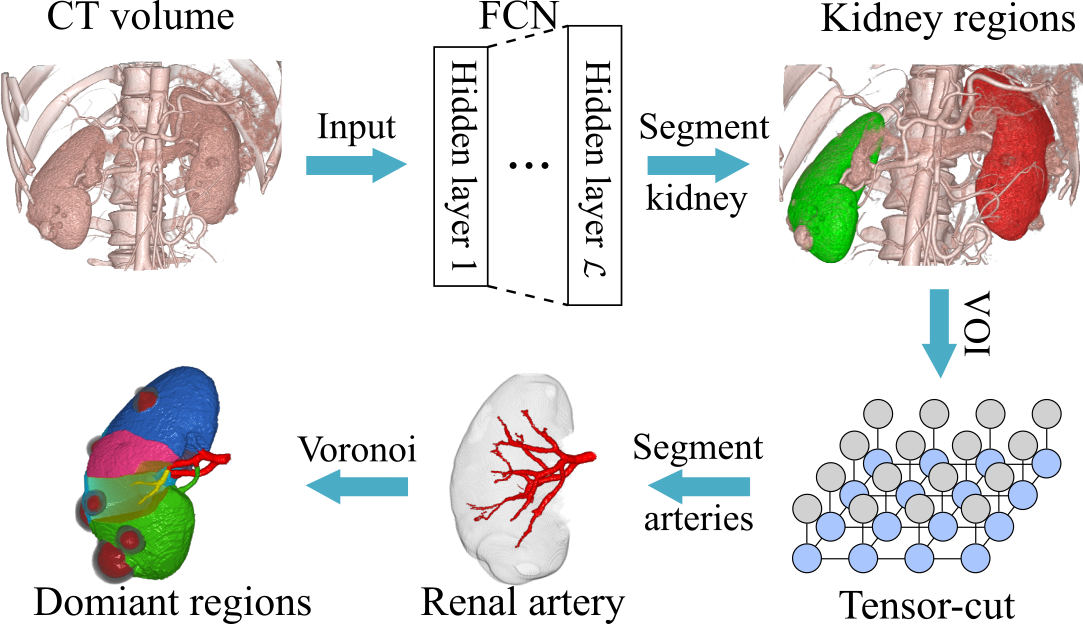}
  \caption{Workflow: Our precise estimation approach can be divided into three parts: kidney segmentation, renal artery segmentation and estimation of vascular dominant regions.}\label{fig:workflow}
\end{figure}

\section{Methods}\label{method}
In this section, we describe our proposed methods in detail. The workflow is shown in Fig. \ref{fig:workflow}. First, the kidney regions are extracted using a deep learning approach. Second, a fine renal artery segmentation method is performed to segment the renal arteries inside the bounding-box of the kidney regions. After extracting the kidneys and the renal arteries, we estimate the vascular dominant regions with a Voronoi diagram. The relative statistics of the dominant regions are calculated for further surgical planning.

\subsection{Kidney segmentation} \label{sec:kidney_segmentation}

In this work, we segment the kidney regions with a 3D U-Net-like fully convolutional network (FCN) architecture. U-Net architecture \cite{ronneberger2015u,cciccek20163d}, which is an extended version of FCN architecture, consists of a contracting path and a symmetric-expanding path. U-Net can achieve high segmentation accuracy with sparse annotated data \cite{cciccek20163d}. Recently, many U-Net-like architectures have been proposed for segmentation tasks \cite{roth2018application,roth2018towards,milletari2016v,shen2018influence,zhu2018anatomynet,zhu2018deeplung,zhu2018deepem,zhu2018adversarial}.

Our network is based on previous 3D U-Net-like architecture \cite{roth2018towards,shen2018influence}. Roth et al. presented a U-Net-like architecture for organ segmentation on 3D medical images and achieved state-of-the-art segmentation results \cite{roth2018towards}. To tackle the GPU memory limitation problem, they used a sliding-windows strategy for large medical data. However, these cropped sub-volumes were trained independently, i.e., the spatial position information of the sub-volumes was ignored during training. Spatial information is a critical feature for organ segmentation because the relative spatial position of the human organs is generally unchanged between patients. Exploiting spatial information should improve organ segmentation accuracy. Many works have involved spatial information into networks. Brust et al. directly incorporated absolute position information into fully connected layers \cite{brust2015convolutional}. Akoury et al. presented a ``Spatial PixelCNN'' to impose spatial prior information to maintain the coherence of generated synthetic images \cite{akoury2017spatial}. Chen et al. incorporated spatial information into the end of an encoder (a bottom feature map) \cite{chen20173d,wolterink2015automatic}. Zhu et al. incorporated spatially structured learning in an adversarial FCN for mammographic mass segmentation \cite{zhu2018adversarial}.

\begin{figure}
  \includegraphics[width=\textwidth]{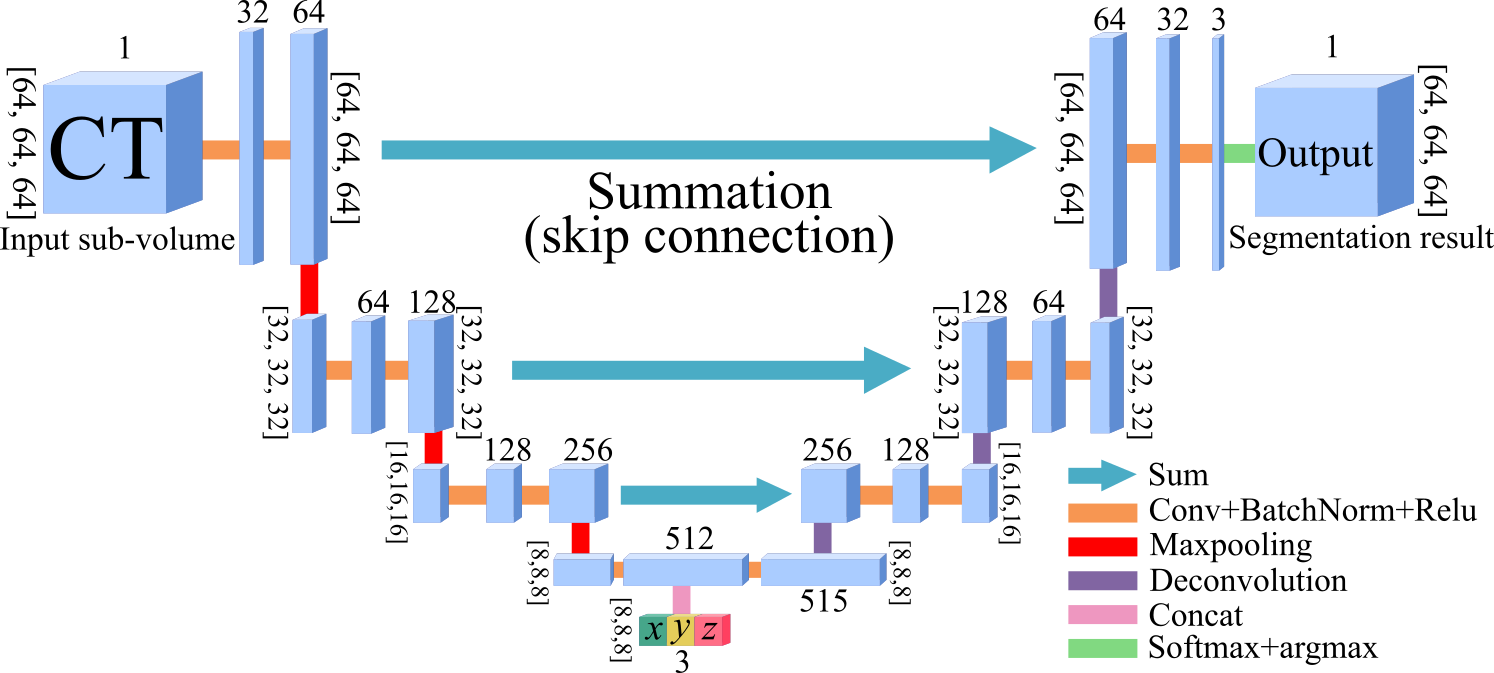}
  \caption{Architecture of our presented FCN.} \label{fig:unet}
\end{figure}

Inspired by the work of Chen et al. \cite{chen20173d}, we introduce spatial position information into 3D U-Net-like architecture to impose the spatial information of each cropped sub-volume into our FCN architecture. Our proposed network is illustrated in Fig. \ref{fig:unet}. The backbone U-Net-like structure, which is based on a previous work \cite{rothjamit2018}, consists of four resolution levels. At each level, a skip connection links the contracting path and the corresponding symmetric-expanding path to provide higher resolution features to the symmetric-expanding path. Unlike original U-Net architecture, the skip connections in this network are summation instead of concatenation. {Summation connections were first incorporated in U-Net by Roth et al. \cite{roth2018application}. Their experimental results show that summation connections are slightly better than the original concatenation connections in the pancreas segmentation task.} Each resolution level contains two series of convolutional layers, batch normalization and ReLU activation, in both the contracting and symmetric-expanding paths. The kernel sizes of all the convolutional and deconvolutional layers in our network are fixed to $3\times 3\times 3$. The kernel size of the max pooling layers is fixed to $2\times 2\times 2$.

We concatenated a three-channel feature map, including $x, y, z$ coordinates, to the bottom feature map to introduce the position information to FCN. The input coordinate information is the relative coordinates of the input sub-volume in the entire CT normalized to $[0, 1]$. Let $\mathbfcal{P}$ be the three-channel spatial feature map, thus $\mathbfcal{P}$ is defined as $\mathbfcal{P} = [\frac{\mathbf{x}}{W}, \frac{\mathbf{y}}{H}, \frac{\mathbf{z}}{D}]$, where $\mathbf{x}$, $\mathbf{y}$, and $\mathbf{z}$ denote coordinates of voxels in a sub-volume. $W, H, \text{ and } D$ denote width, height and depth of a CT image. Unlike a previous study \cite{chen20173d} that only considered the center coordinates of the input sub-volume, we used all of the position information and resized the position volume (containing coordinate information) to a suitable input size. 

\begin{figure}[tb!]
  \centering
  \newcommand{\deffigure}[1]{%
    \includegraphics[width=0.4\linewidth,trim={0cm 12.5cm 26cm 12.5cm},clip]{#1}}
   \newcommand{\deffigureA}[1]{%
    \includegraphics[width=0.4\linewidth,trim={0cm 9.3cm 18.5cm 8.5cm},clip]{#1}}
  \subfloat[Original sub-volume]{\deffigure{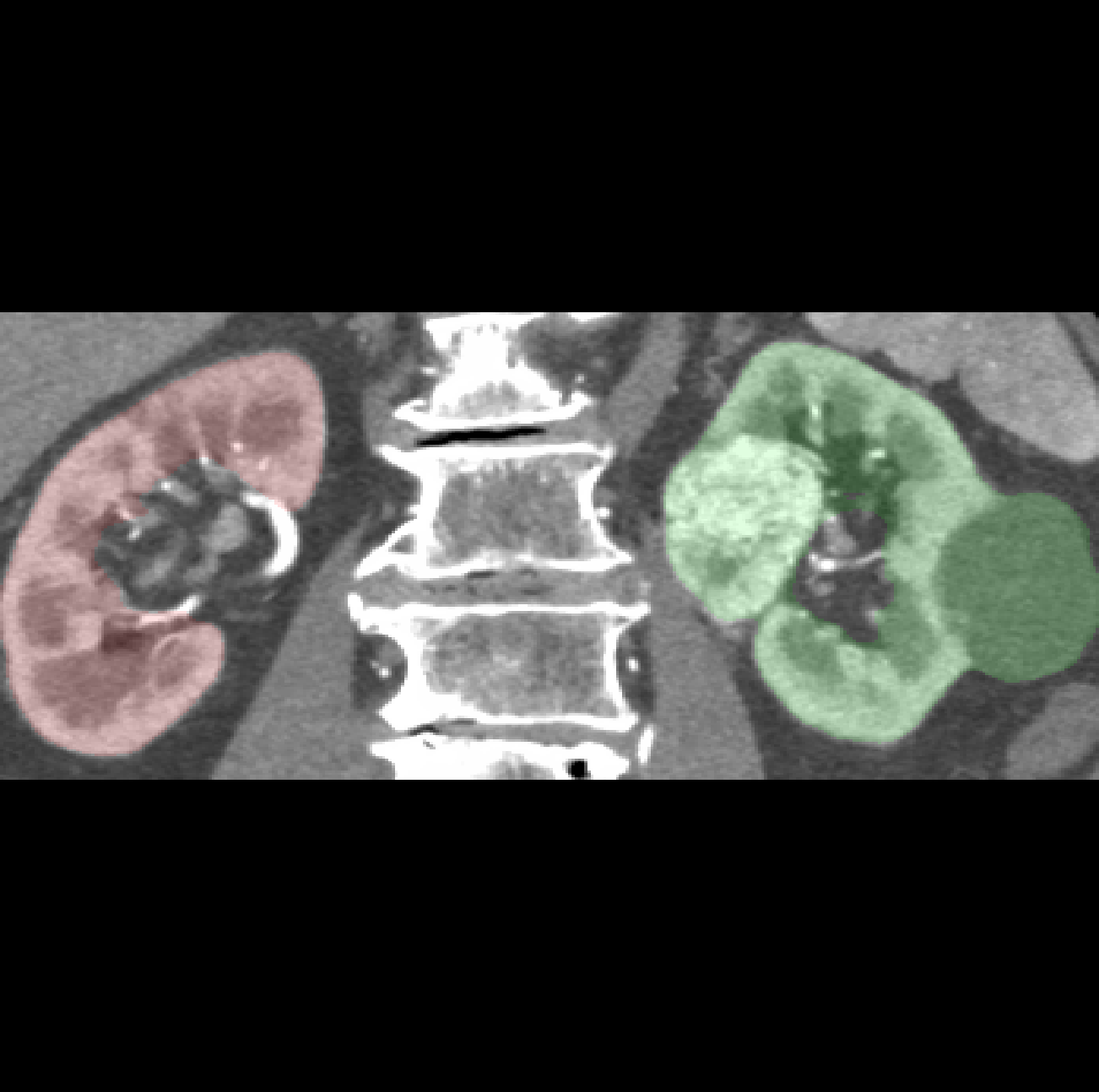}}
  \subfloat[Rigid transformation]{\deffigure{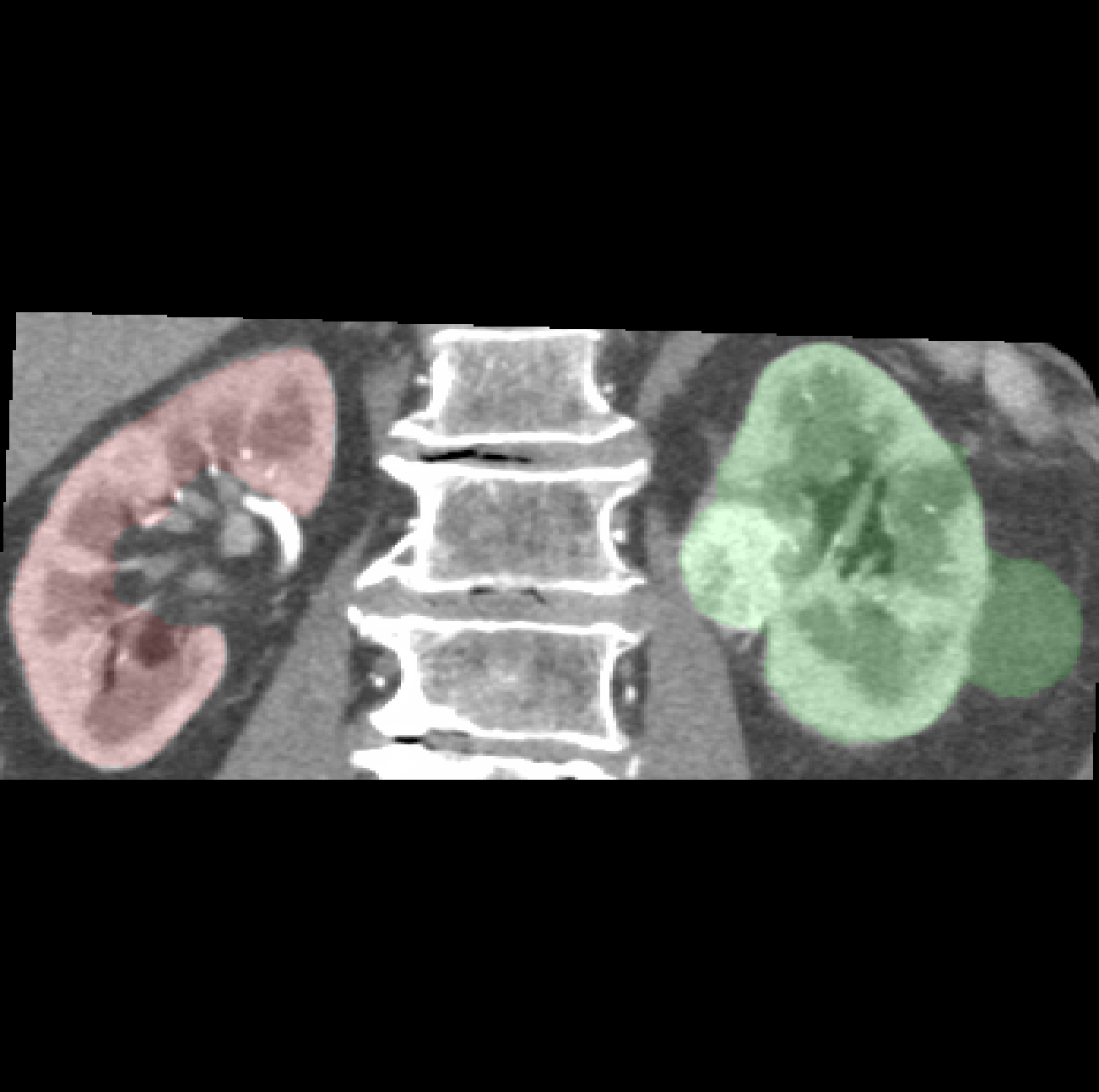}} \\
  \subfloat[Elastic transformation]{\deffigure{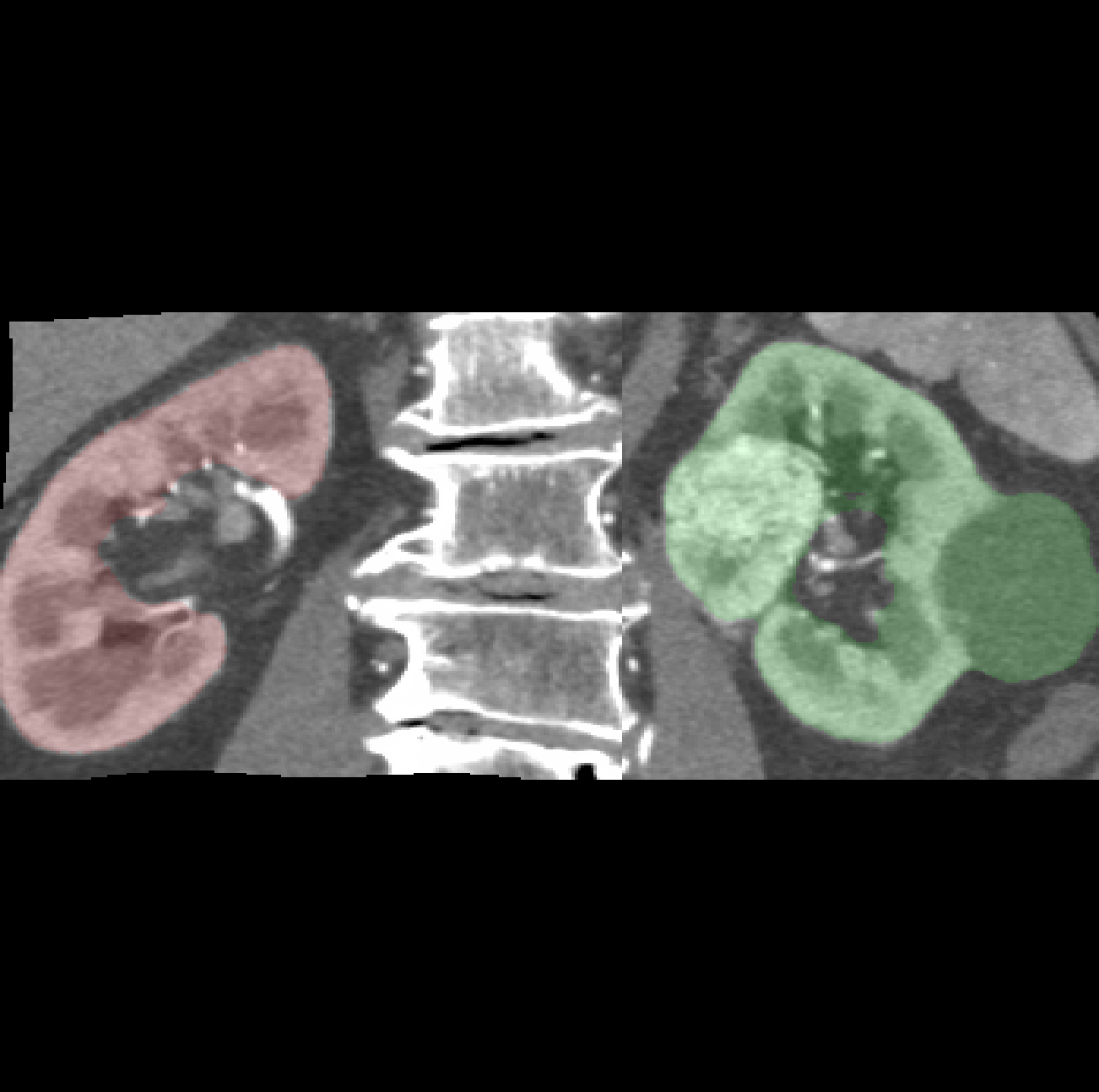}}
  \subfloat[Hybrid transformation]{\deffigureA{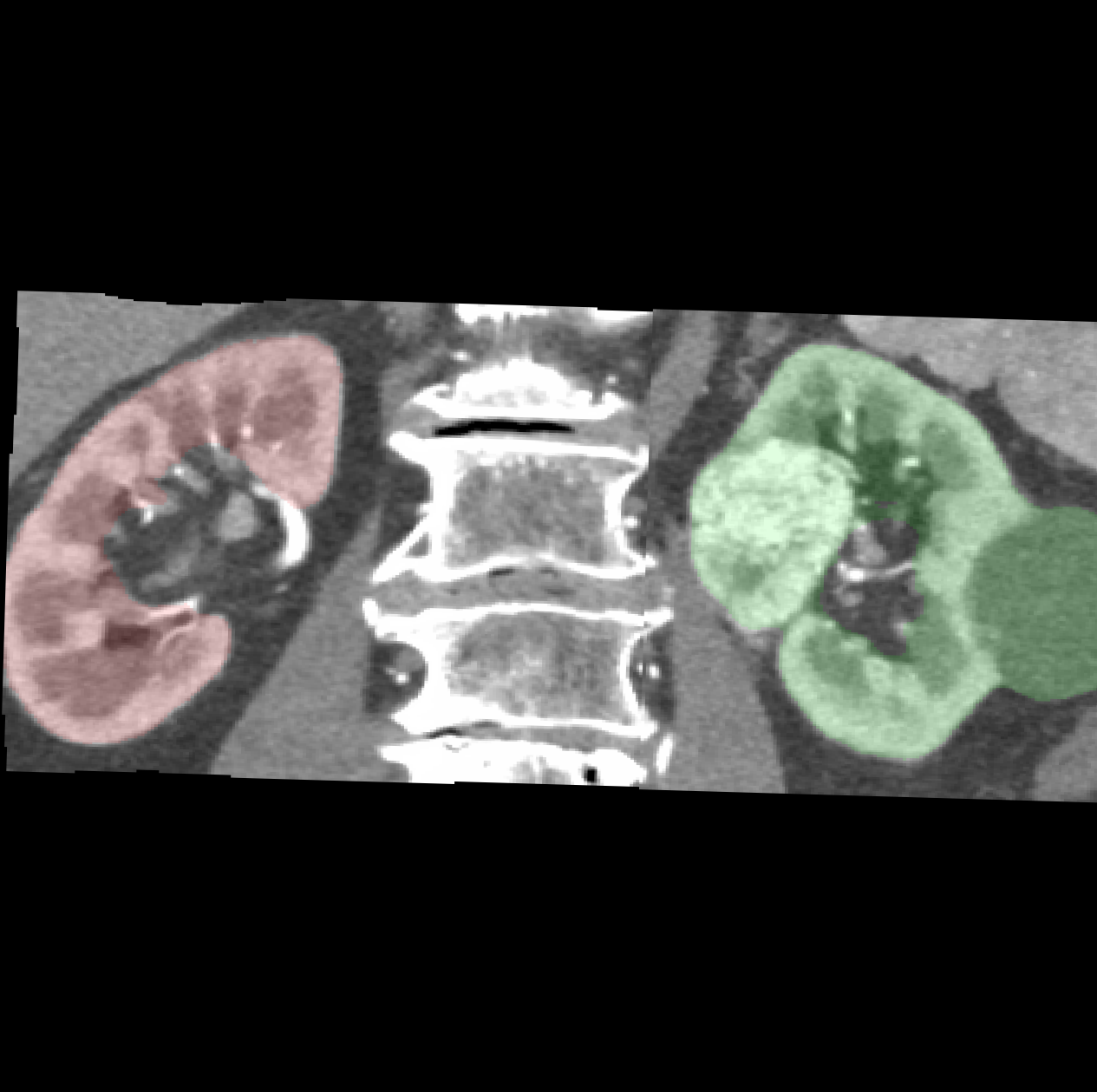}}
  \caption{Examples of data augmentation: (a) original sub-volume extracted from CT data. Red region indicates ground-truth label of kidney. (b) and (c) rigid and elastic transformation results. (d) transformation result containing both rigid and elastic transformations}\label{fig:aug}
\end{figure}

\subsubsection*{Training}
In this work, the input volume size of our network is fixed to $N_x\times N_y\times N_z$. At each epoch, $n$ sub-volumes are cropped from the original CT volume and fed to the neural network. Here $n$ denotes the batch size. To achieve the best segmentation performance, we exploit the transfer learning technique and pre-train our model on a multi-organ segmentation dataset \cite{roth2018application,shen2018influence}, which doesn't contain any kidney annotation, and fine-tune the model on our kidney dataset. {This multi-organ segmentation dataset contains 377 cases, with 340 cases used for pre-training and 37 cases used for validation. The model with the best validation performance is used for fine-tuning. We used all pre-trained layers except the last classification layer. Fine-tuning was done on all layers.}

Similar to previous works \cite{ronneberger2015u,cciccek20163d,roth2018application,roth2018towards,shen2018influence}, we used a data-augmentation technique to increase the data variety and robustness. We performed both rigid and elastic transformations to each cropped sub-volume. Rigid transformation includes a translation with a range of $[-10, +10]$ pixels at each dimension and a rotation with range of $[-15^{\circ}, +15^{\circ}]$. A B-spline deformation is employed as an elastic transformation to each sub-volume, like in previous works \cite{cciccek20163d,roth2018application,roth2018towards}. {For each sub-volume, the deformation fields are randomly sampled from a uniform distribution with a maximum displacement of 3, and the number of B-spline control points in each dimension is set to 3 for all data-augmented experiments.} In this work, we performed an in-place deformation operation. In a single iteration, we fed one original sub-volume and $n-1$ hybrid deformed sub-volumes to the network.
A data-augmentation example is shown in Fig. \ref{fig:aug}. We individually show the rigid transformation, the elastic transformation, and the hybrid transformation, which combines the rigid and elastic transformations. Deformation is computed on-the-fly at each epoch during training.

In this work, we use pseudo Dice loss instead of conventional cross-entropy loss. Dice loss was first introduced in 2016 \cite{milletari2016v}. The definition of multi-class Dice coefficient loss $D$ used in this work can be written:
\begin{equation}
D=-\frac{1}{K}\sum_{k=1}^{K}\left(\frac{2\sum_{i}^{N} p_{i,k}g_{i,k}}{\sum_{i}^{N} p_{i,k}^2+\sum_{i}^{N} g_{i,k}^2}\right).
\label{equ:dice_loss_multi}
\end{equation}
where $g_{i,k} \in G$ and $p_{i,k} \in P$ denote voxels from the ground-truth volume and segmentation results for class $k$ of total $K$ classes. $N$ is the total voxel amount of the volume.

\subsubsection*{Testing}
The input volume size for testing is the same as in the training. We used a sliding-window strategy to obtain sub-volumes with size $N_x\times N_y\times N_z$ for testing. After predicting all of the sub-volumes, we restored these predictions to a complete 3D labeling data based on their respective positions. The probabilities of the overlapping regions were computed as average probabilities: $p(x) = \frac{1}{R}\sum^R_r p_r(x)$, where $p_r(x)$ denotes the probability of voxel $x$ from $r$-th sub-volume, $r\in R$. 

\subsection{Renal artery segmentation} \label{sec:vessel_segmentation}

After extracting the kidney region from a CT volume, we performed a renal artery segmentation inside the VOI of the kidney region. Our previous work presented a tensor-based graph-cut blood vessel method called tensor-cut to segment the renal arteries \cite{wang2016tensor}. This blood vessel segmentation method can segment tiny vascular structures, primarily designed for renal artery segmentation. Tensor-cut models the tubular structure as a second-order tensor using a Hessian matrix and builds a first-order Markov random field (MRF) using both geometry (tensors) and intensity information. Finally, a graph-cut approach is utilized to find the optimal MRF solution.

\iffalse
The graph-cut approach is widely used in the computer vision field for many low-level image processing problems \cite{}. A max-flow min-cut algorithm is used to find the global minimum solution of the given MRF. 

To reduce the manual labor of creating the initial terminal nodes for a graph-cut, we utilized a graph-cut method to extract the thick blood vessels. Convolutional graph-cut method need both foreground and background voxels specified manually. But for tubular structure segmentation problem, more foreground information is expected to obtain a better segmentation result. Besides, the small blood vessels make it hard to specify foreground voxels manually. In this work, we utilize vesselness enhancement filter to extract the rough blood vessels as foreground information. \fi 

\begin{figure}[tb]
  \centering
  \includegraphics[height=4.5cm]{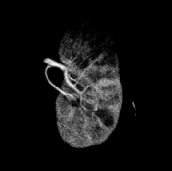}
  \includegraphics[height=4.5cm]{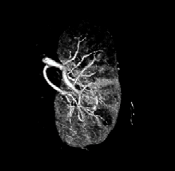}
  \caption{Left: VOI of original 3-D CT volume of kidney region $I(\textbf{x})$. Right: result of vesselness enhancement filter $I_{\mathcal V}(\textbf{x})$.}
  \label{fig:hessian}
\end{figure}{}

\subsubsection*{Vesselness enhancement filter}
Vesselness enhancement filters have been widely used in the image processing field since they were proposed in 1998 by Frangi et al. \cite{frangi} and Sato et al. \cite{sato}. By analyzing the eigenvalues of a Hessian matrix, these filters can obtain high response against contrast changes. The Hessian matrix of a 3-D image $I(\textbf{x})$ is given by 
\begin{equation}
\nabla^2 I(\textbf{x}) = 
\begin{bmatrix}
I_{xx}(\textbf{x}) &I_{xy}(\textbf{x}) &I_{xz}(\textbf{x}) \\
I_{yx}(\textbf{x}) &I_{yy}(\textbf{x}) &I_{yz}(\textbf{x}) \\
I_{zx}(\textbf{x}) &I_{zy}(\textbf{x}) &I_{zz}(\textbf{x}) \\
\end{bmatrix},
\end{equation}
where $I_{xx}(\textbf{x}) = \dfrac{\partial^2}{\partial x^2}I(\textbf{x})$ stands for the second-order partial derivatives of image $I(\textbf{x})$ around voxel $\textbf{x}$. The following is Sato's quantitative vesselness measure, which is specialized to 3-D images where the vessels are brighter than the background: 
\begin{equation}
\begin{aligned}
\mathcal V = \begin{cases}
|\lambda_3| \left( \dfrac{\lambda_2}{\lambda_3} \right) ^{\gamma^{23}} \left( 1+\dfrac{\lambda_1}{|\lambda_2|} \right) ^{\gamma^{12}}, & \text{if } \lambda_3 < \lambda_2 < \lambda_1 \leq 0 \\
0, & \text{otherwise.}
\end{cases}
\end{aligned}
\end{equation}
Utilizing a quantitative vesselness measure, we obtain the result of vesselness enhancement filter $I_{\mathcal V}(\textbf{x})$ (Fig.~\ref{fig:hessian}). The enhanced vessel radius ranges from 1 to 2 mm. 

A Hessian matrix can be treated as a second-order tensor. The vesselness enhancement filter transfers the higher dimensional tensor to 1-dimensional Euclidean measurement $\mathcal V$. This transformation introduces external errors. To tackle this problem, our tensor-cut algorithm models the Hessian matrix as a tensor and uses a Riemannian metric to measure the tensors.

%Next, a simple clustering method, k-means, is applied to extract the definite blood vessels from $I_{\mathcal V}(\textbf{x})$. We set clustering sets number to 4 as prior knowledge. Only the intensity of image data $I_{\mathcal V}(\textbf{x})$ is clustered, component with the maximal mean intensity will be selected as foreground object. Fig.~\ref{fig:graph-cut} shows the result of clustering. 

\begin{figure}[tb]
  \includegraphics[width=\linewidth]{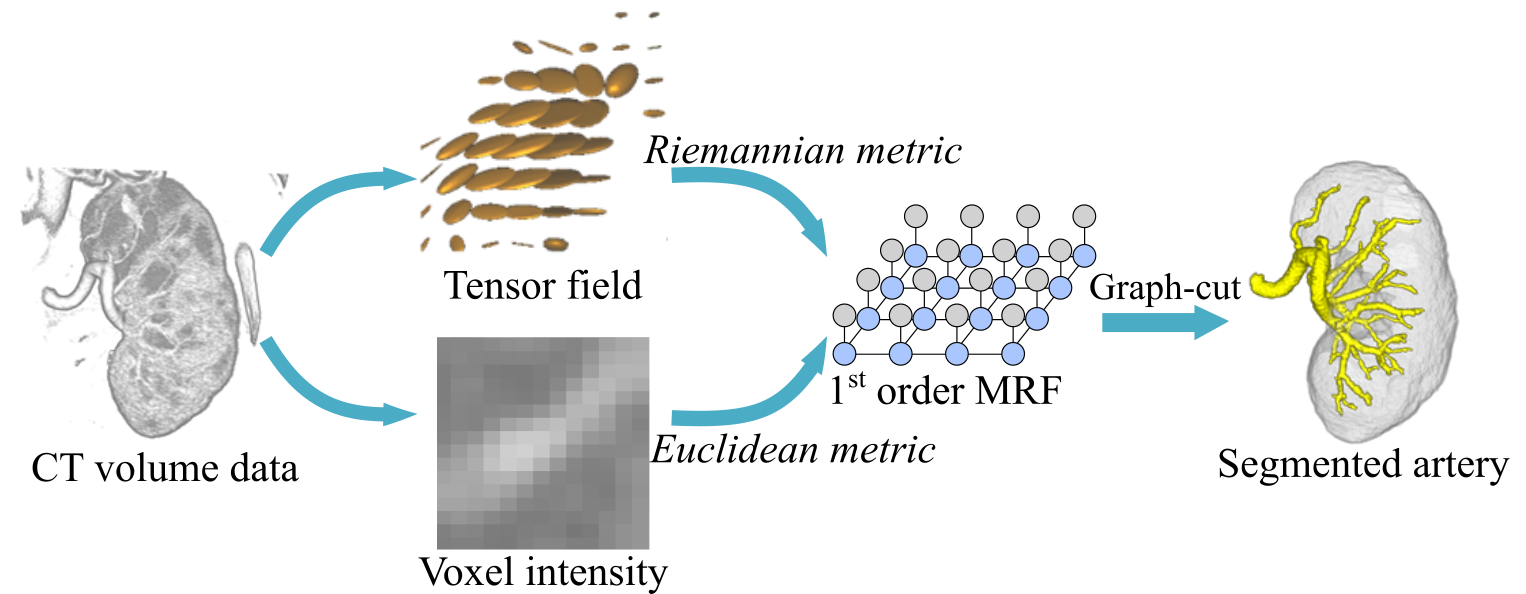}
  \caption{The tensor-cut workflow: Both tensors and voxels are used to create a first-order Markov random field (MRF). Then graph-cut algorithm finds optimal solution of the given MRF that corresponds to the final segmentation result. }
  \label{fig:tensorcut}
\end{figure}

\subsubsection*{Tensor-cut}

The main idea of the tensor-cut method is to build a first-order MRF using both the geometry and intensity information and a graph-cut algorithm to extract the vascular structures. A simple tensor-cut workflow of tensor-cut is illustrated in Fig. \ref{fig:tensorcut}. In this section, we briefly introduce the tensor-cut algorithm. More details are available \cite{wang2016tensor}.

As mentioned above, Hessian matrix $\nabla^2 I(x_i)$ $(x_i \in\textbf{x})$ can be treated as a second-order tensor $\mathcal T$. First, a vesselness enhancement filter is applied to the CT volume to obtain tensor field $\mathbb T = \{\mathcal T_1, ..., \mathcal T_n\}, n\in N$, where $N$ denotes the total voxel number.

To build a MRF with tensors, such tensor characteristics are needed as distance and mean value. We calculated them with an Euclidean metric \cite{wang2016precise}. Tensors resemble ellipsoids. The dissimilarity is calculated by the length of three principal axes, the angle, and the distance between center points. However, the Euclidean metric ignores the property by which tensors should lie on a manifold space. The tensor-cut method uses a Riemannian metric to measure the distance between two tensors. We used the affine invariant Riemannian metric presented by Pennec et al. \cite{pennec2006riemannian}. A detailed mathematical proof is available \cite{moakher2005differential,pennec2006riemannian}.

\iffalse
An affine invariant Riemannian metric for symmetric positive definite (SPD) matrices and mean value of the  arbitrary matrix space were presented \cite{pennec2006riemannian,moakher2005differential}. 
The geodesic distance between two tensors is defined:
\begin{equation}
  d(\mathcal T_1, \mathcal T_2) = \left( \sum_{i=1}^d\log ( \lambda_i(\mathcal T_1, \mathcal T_2) )^2 \right)^{1/2},
\end{equation}
where $\lambda(\mathcal T_1, \mathcal T_2)$ denotes the generalized eigenvalues of tensors $\mathcal T_1$ and $\mathcal T_2$. $d$ is the order of the SPD matrix, and here the $d$ of the Hessian matrix is 3.

Based on the Fr\'echet mean algorithm, the mean value of the arbitrary matrix space can be obtained by minimizing the Fr\'echet variance: $\bar{\mathcal T} =\argmin\sum_{i=1}^N \allowbreak d^2(\mathcal T_i, \mathcal T)$. The Newton gradient descent is used to solve this optimization problem. Here, we give the mean tensor at $t+1$ steps: 
\begin{equation}
  \bar\T_{t+1} = \bar\T^{\frac{1}{2}} \exp\left( \frac{1}{N} \sum_{i=1}^N \log (\bar\T_t^{-\frac{1}{2}} \T_i \bar\T_t^{-\frac{1}{2}} )\right)\bar\T_t^{\frac{1}{2}}.
\end{equation}
This iteration begins with arbitrary tensor $\bar\T_0\in\mathbb{T}$. A detailed mathematical proof is available \cite{pennec2006riemannian}.\fi

Instead of using a conventional histogram probability estimation approach, we used a Gaussian mixture model (GMM) \cite{rother2004grabcut}. Two GMMs are required for presenting the distribution of the foreground and background regions. GMM distribution under label $\mathbf{L} = \{L_F, L_B\}$ is denoted by $\text{Pr}(\mathbf{x}|\mathbf{L})$, $L_F$ and $L_B$ are the foreground and background labels. Thus, MRF's energy function can be defined:
\begin{equation}
  \begin{aligned}
    E(L) &= \underbrace{\sum_{\mathbf{x}\in\mathbb{X}}-\log \Pr(\mathbf{x}|L_{\mathbf{x}}) + \lambda_I\sum_{\{\mathbf{x}_m, \mathbf{x}_n\}\in\mathcal{N}}V_{m,n}(\mathbf{x}_m, \mathbf{x}_n)}_{\text{intensity term}}\\
    &+ \underbrace{ \omega \biggl(\sum_{\T\in\mathbb{T}}-\log \Pr(\T|L_{\T}) + \lambda_T{\sum_{\{\T_m, \T_n\}\in\mathcal{N}'}}U_{m,n}(\T_m, \T_n)}_{\text{tensor term}}\biggr),\\
    \end{aligned}
\end{equation}
where $\lambda_I$, $\lambda_T$, and $\omega$ are constant parameters to adjust the weight between the two smoothness terms and the tensor term. Two smoothness terms, $V_{m,n}(\cdot, \cdot)$ and $U_{m,n}(\cdot, \cdot)$, present the dissimilarity of the intensity and tensor terms. As in previous works \cite{rother2004grabcut,boykov06,boykov}, we used the Potts model. Obviously, the biggest difference between the conventional graph-cut energy and tensor-cut energy functions is the introduction of the tensor term. By introducing an external tensor term, a more accurate MRF model can be obtained for tubular structure segmentation.

\subsection{Estimation of vascular dominant region}\label{sec:voronoi}
We estimated the renal vascular dominant regions with a Voronoi diagram that is widely used for calculating vascular dominant regions \cite{ukimura2012three,komai2014novel,isotani2015feasibility}. Considering the capillaries along the arteries, each branch of the renal arteries is treated as a set of seed points of a Voronoi diagram instead of using the end points of arteries. Let $B_i\in\mathbf{B}, (i \in \mathbb R)$ be a branch of renal arteries, and define Voronoi cell $\mathcal C_i$ function:
\begin{equation}\label{eq:voronoi}
\mathcal C_i = \lbrace \mathbf{x}\in\mathbf{X}_\mathbf{v} | d(\mathbf{x}, B_i) \leq d(\mathbf{x}, B_j)\rbrace, \text{for all } i\neq j,
\end{equation}
where image voxels $\mathbf{x}$ are inside of kidney region $\mathbf{X}_\mathbf{v}$ extracted by the FCN approach. $d(\cdot)$ is the Euclidean distance between two points, and $d(\mathbf{x}, B_i)$ denotes the minimal Euclidean distance from point $\mathbf{x}$ to vessel branch $B_i$. This is a simple simulation of a real cell getting nutrition from blood vessels.

\begin{figure}[tb]
\centering
\includegraphics[width=\linewidth]{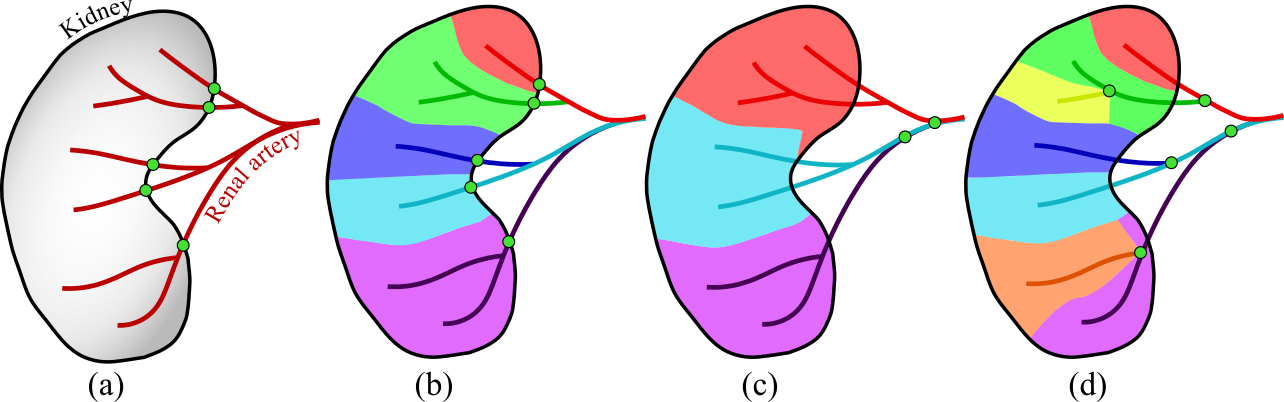}
\caption{Different Voronoi partition results depending on different branch clustering: (a) a simple kidney and renal arteries, where green dots denote entries of renal arteries into the kidney. (b), (c), and (d) Voronoi partition results of three different branch clustering strategies. }\label{fig:voronoi_illus}
\end{figure}

As demonstrated in Eq. \ref{eq:voronoi}, the main factors affecting the accuracy of a Voronoi diagram are kidney region $\mathbf{X}_\mathbf{v}$ and renal artery branches $\mathbf{B}$. In this work, we also provide different partition levels to help physicians better grasp the relationships among arteries, dominant regions, and tumors. Fig. \ref{fig:voronoi_illus} simply illustrates the Voronoi partition strategy used in this work. Fig. \ref{fig:voronoi_illus}(a) shows the kidney region and its renal arteries. Green dots represent the entries of the renal arteries in the kidney. Fig. \ref{fig:voronoi_illus}(b) shows the branch cluster result based on the entries of the arteries. All the artery branches downstream of the entries are clustered into the same group. This clustering idea is treated as a basic rule. Based on it, we generate different Voronoi partition levels by moving the bifurcation level of the vascular tree. As shown in Fig. \ref{fig:voronoi_illus}(c), by moving the bifurcations of Fig. \ref{fig:voronoi_illus}(b) one level upstream, we get new Voronoi partition results that show a coarser partition result. Moving the bifurcations one level downstream leads to a more precise partition result (Fig. \ref{fig:voronoi_illus}(d)).

To get a quantitative measure for the estimation of the vascular dominant regions, we calculated the volume and the volume ratio of each dominant region denoted by $Vol$ and $R$. If dominant regions are adjacent to a tumor, we also calculated the contact area of each adjacent region.

\newcommand{\caseNum}{27 }
\section{Materials}\label{materials}
In this work, we used \caseNum pieces of abdominal contrast-enhanced CT volume data to evaluate the performance of our proposed method. The pixel spacing ranged from $0.665$ to $0.742$ mm, and slice pitch ranged from $0.4$ to $2.0$ mm. All \caseNum cases containing 54 kidneys were used to evaluate the accuracy of the kidney segmentation. Eight cases containing 14 kidneys were used for evaluating the dominant-region estimation. The ground truth of the kidneys was created by an engineer with medical knowledge. The ground truth of the renal artery was created by two engineers with medical knowledge. Since we directly used our previous method for blood vessel segmentation, we did not perform a quantitative evaluation for blood vessel segmentation in this work. Detailed experimental results are available in our previous work \cite{wang2016tensor}.

\begin{figure}[tb]
\centering
\includegraphics[width=0.8\linewidth]{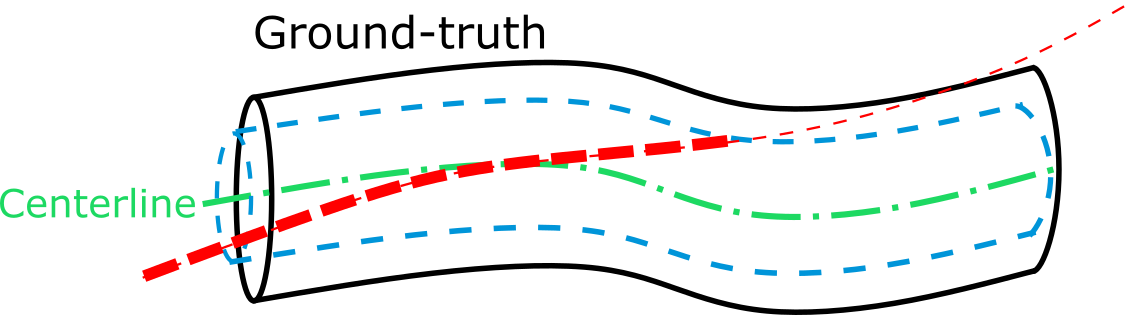}
\caption{Illustration of computing centerline overlap. Tubular structure in black is gold standard blood vessel, and its centerline $\Omega_G(\textbf{x})$ is shown as green dash-dot line. Tube $\mathcal U(\textbf{x})$ is generated by dilating the centerline of radius 1 voxel shown in blue. Red dashed line represents centerline of segmented vessels $\Omega_O(\textbf{x})$. Overlapped centerline $OV = \lbrace \textbf{x} | \textbf{x}\in\Omega_O(\textbf{x}), \textbf{x}\in\mathcal U(\textbf{x}) \rbrace$. $CO = 2 * \lVert OV\rVert / (\lVert\Omega_G\rVert + \lVert\Omega_O\rVert)$, where $\lVert\cdot\rVert$ represents length.}
\label{fig:centerline}
\end{figure}

\begin{figure}[tb]
\centering
\subfloat[Pre-trained]{\includegraphics[width=0.48\linewidth]{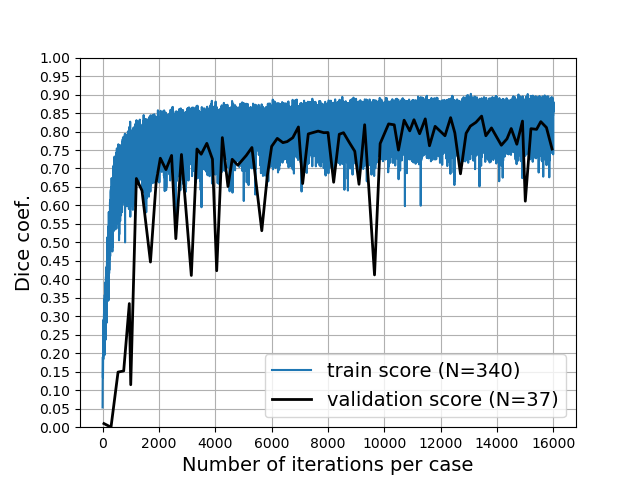}}
\subfloat[Fine-tuned]{\includegraphics[width=0.48\linewidth]{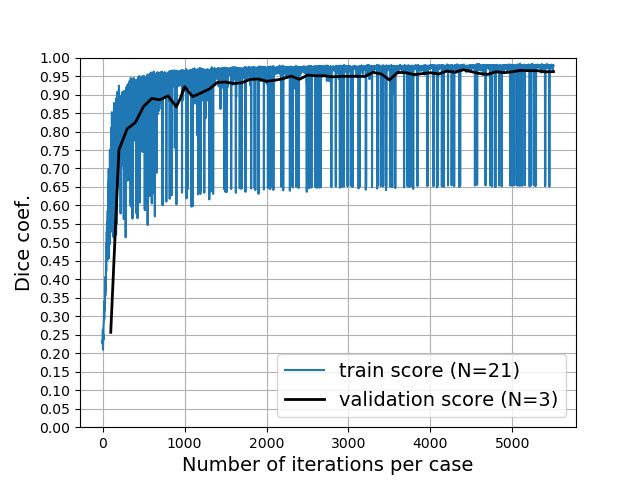}}
\caption{Pre-training and fine-tuning learning curves: Blue curve denotes training Dice coefficient ($DSC$), and black curve denotes {validation} $DSC$. Maximum {validation} $DSC$ of pre-training achieved $88.8\%$, and maximum {validation} $DSC$ of kidney segmentation achieved $96.7\%$.}
\label{fig:learningcurve}
\end{figure}

\section{Experiments and Results}

\subsection{Kidney segmentation}

We used an 8-fold cross-validation scheme to evaluate the accuracy of kidney segmentation. {We divided all \caseNum CT volume data into train/validation/test splits at a ratio of $0.8/0.1/0.1$. The model with the best validation performance is to be used for test.} For a quantitative evaluation, we used three measures: the Dice Similarity coefficient ($DSC$), Sensitivity ($Se$), and the Hausdorff distance ($HD$). {$DSC$ is a commonly used measure in image segmentation. It is able to reflect the general segmentation ability. $Se$, also known as recall rate, measures the ability of the method to extract correct kidney regions. $HD$ is introduced to measure the surface distance between segmentation results and ground truth. We use the $HD$ metric to focus on segmentation accuracy of the kidney itself. Therefore, as post-processing, we extracted the top 2 largest connected components as kidney regions. $HD$ is used to measure the post-processed segmentation results, to reflect the actual segmentation ability of end-to-end FCN, $DSC$ and $Se$ are still used to measured the original segmentation results without any post-processing.} $DSC$, $Se$, and $HD$  are defined as follows:
\begin{equation}
    \begin{aligned}
      DSC &= \frac{2*TP}{2*TP+FP+FN}, Se = \frac{TP}{TP+FN}, \\
      HD &= \max(h(S_{gt},S_{seg}), h(S_{seg}, S_{gt})),
    \end{aligned}
\end{equation}
where True Positive (TP), False Positive (FP), True Negative (TN), and False Negative (FN) were measured in a voxel-wise way. $h(S_{gt}, S_{seg}) = \max_{p_{gt}\in S_{gt}}\min_{ p_{seg}\in S_{seg} }\allowbreak\lVert p_{gt} - p_{seg} \rVert$, where $p$ denotes the voxel coordinate that belongs to surface $S$.  $S_{gt}$ and $S_{seg}$ represent the surfaces of the manually annotated ground-truth label and post-processed segmentation results. 

Experiments were performed on an NVIDIA Quadro P6000 with 24 GB memory. Training on {21} cases took about $7-8$ hours for 4000 iterations, and the testing phase took about five minutes for a single case.  For all experiments in this work, we set the learning rate to $0.01$, the batch size {$n$} to 6, {the sub-volume size $N_x\times N_y\times N_z$ to $64\times 64\times 64$}, and the epoch number to 4000 {for fine-tuning and 16000 for pre-training}.

\begin{figure}
    \centering
    \includegraphics[width=0.9\linewidth]{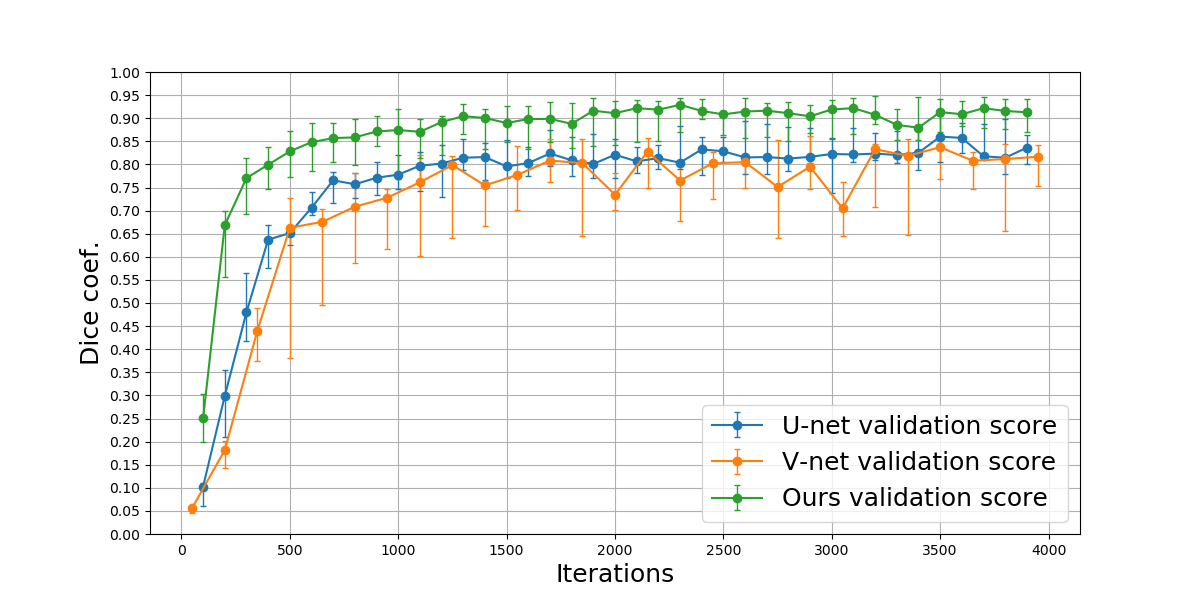}
    \caption{Validation $DSC$ of U-Net \cite{roth2018towards}, V-Net \cite{milletari2016v}, and our proposed spatially aware FCN. All three of these networks were pre-trained on a multi-organ dataset. Points on curves denote median values of all 8-fold cross validations. Upper and lower bounds of error bars denote the first and third quartiles.}
    \label{fig:validation_curve}
\end{figure}

\begin{figure}[b]
\centering
\newcommand{\smallwidth}{0.11\linewidth}
\newcommand{\vr}[1]{\includegraphics[width=\linewidth]{#1}}
%%%%%%%%%%%%%%%%
\newsavebox{\cfboxbox}
\newcommand{\cfbox}[2]{%
  \mbox{%
    \sbox{\cfboxbox}{#1}%
    \setlength{\fboxsep}{0pt}% don't add space
    \setlength{\fboxrule}{1pt}%
    \color{#2}%
    \fbox{\usebox{\cfboxbox}}%
  }%
}

\begin{tabular}{p{3.2cm}@{}c@{}c@{}c@{}p{3.1cm}@{}c@{}c@{}}
\multirow{2}{*}[2.8em]{\vr{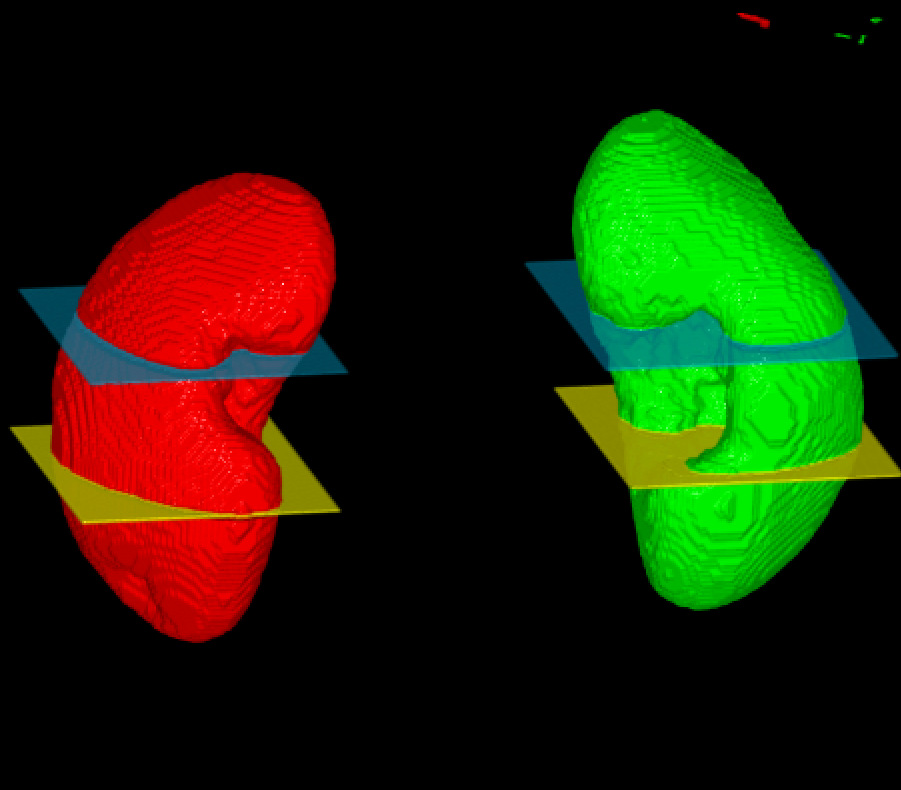}}
&\cfbox{\includegraphics[width=\smallwidth,trim={4cm 6.5cm 12cm 10cm},clip]{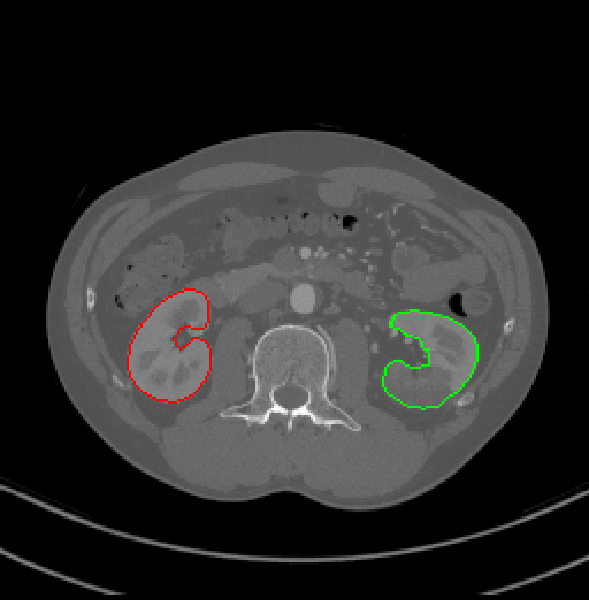}}{yellow}
&\cfbox{\includegraphics[width=\smallwidth,trim={13cm 6cm 3cm 10.5cm},clip]{figures/209kidney_slice93.png}}{yellow}
&
&\multirow{2}{*}[2.8em]{\vr{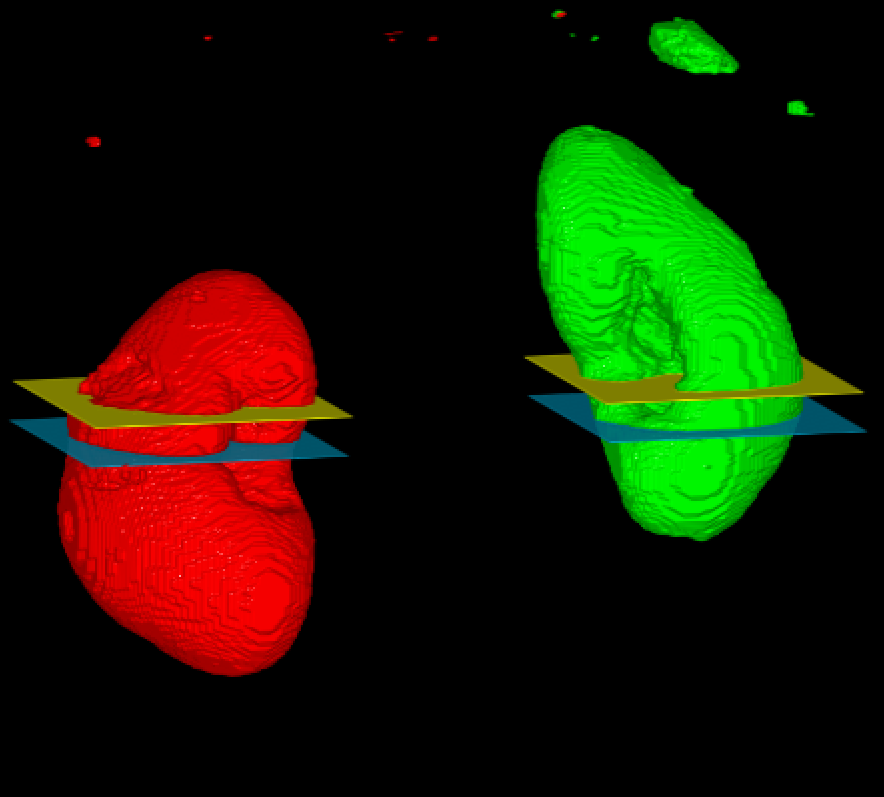}}
&\cfbox{\includegraphics[width=\smallwidth,trim={5cm 7.5cm 13cm 10.5cm},clip]{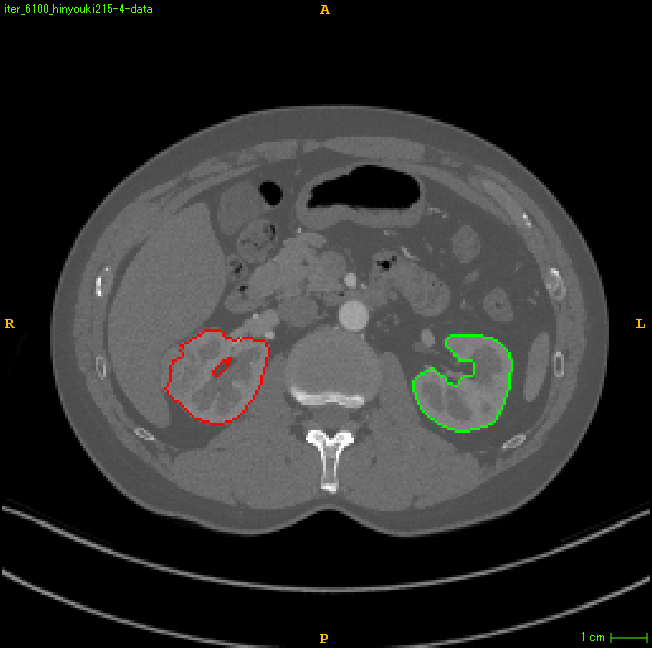}}{yellow}
&\cfbox{\includegraphics[width=\smallwidth,trim={14cm 7cm 4cm 11cm},clip]{figures/215kidney_slice143.png}}{yellow} \\

&\cfbox{\includegraphics[width=\smallwidth,trim={4cm 6cm 12cm 10.5cm},clip]{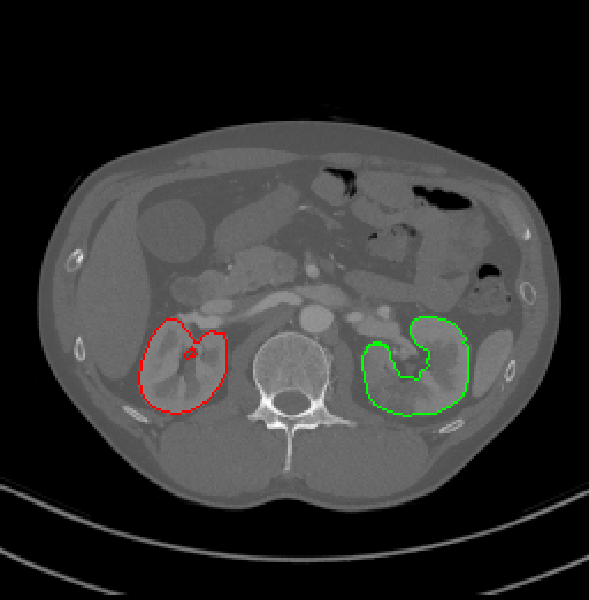}}{blue}
&\cfbox{\includegraphics[width=\smallwidth,trim={12.5cm 6cm 3.5cm 10.5cm},clip]{figures/209kidney_slice73.png}}{blue}
& &  
&\cfbox{\includegraphics[width=\smallwidth,trim={5cm 7.5cm 13cm 10.5cm},clip]{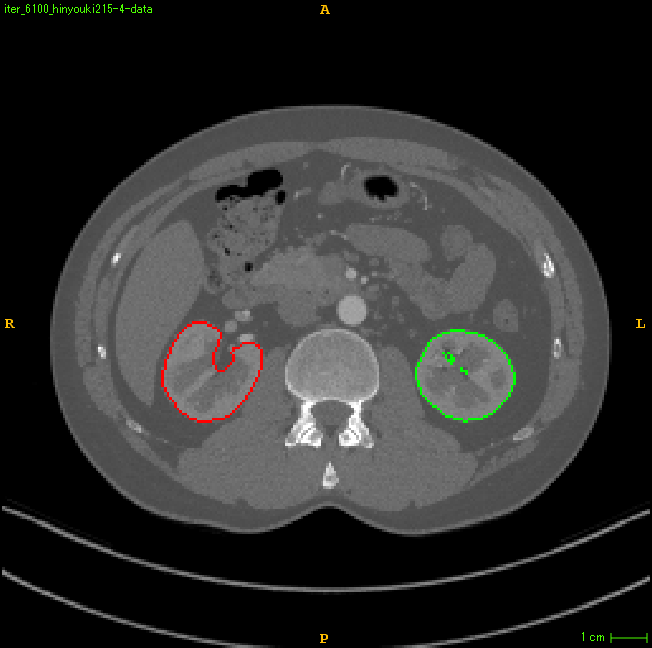}}{blue}
&\cfbox{\includegraphics[width=\smallwidth,trim={14cm 7cm 4cm 11cm},clip]{figures/215kidney_slice157.png}}{blue} \\ 
\multicolumn{1}{c}{Volume rendering}         
& \multicolumn{2}{c}{Axial slices} 
&
& \multicolumn{1}{c}{Volume rendering}
& \multicolumn{2}{c}{Axial slices}
\end{tabular}
\caption{Two kidney segmentation examples using proposed FCN. Both 3D volume rendering and 2D segmented kidneys ROIs are shown. Red and green lines indicate contour lines of segmented kidneys. Yellow and blue sections shown in volume rendering correspond to 2D ROIs with same colors.}\label{fig:kidney_seg}
\end{figure}

\begin{figure}[tb]
    \newcommand{\figA}[1]{\includegraphics[width=0.48\linewidth,trim={0.0cm 0 0.5cm 0},clip]{#1}}
    \newcommand{\figB}[1]{\includegraphics[width=0.48\linewidth,trim={0.2cm 0 0.3cm 0},clip]{#1}}
    \centering
    \begin{minipage}[b]{0.45\linewidth}
    \subfloat[Ground truth]{\figA{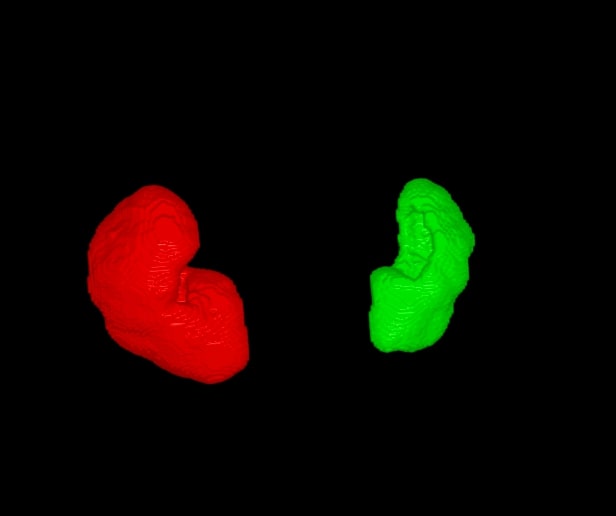}}
    \subfloat[U-Net]{\figA{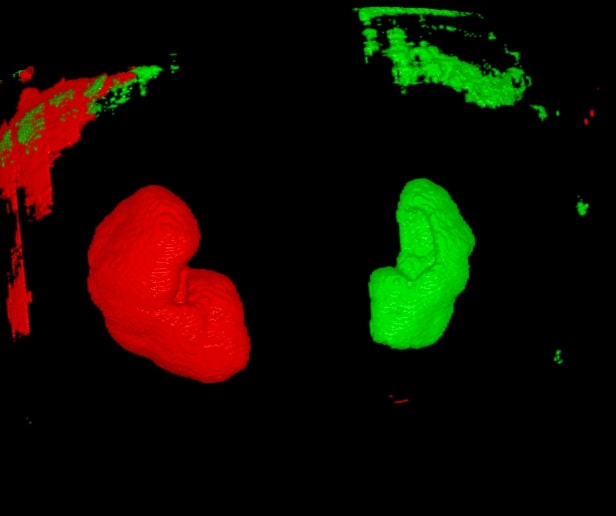}}\\[-1em]
    \subfloat[V-Net]{\figA{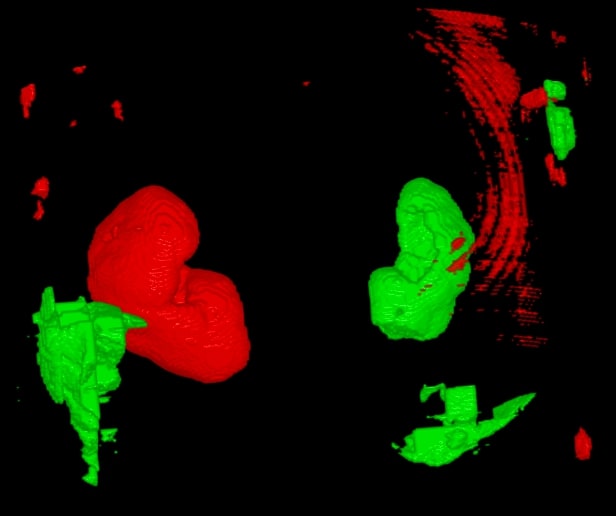}}
    \subfloat[Our proposed]{\figA{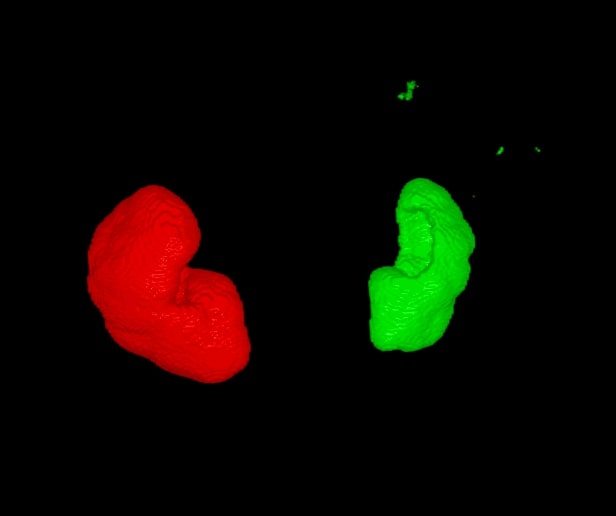}}
    \end{minipage}
    \setcounter{subfigure}{0}
    \begin{minipage}[b]{0.45\linewidth}
    \subfloat[Ground truth]{\figB{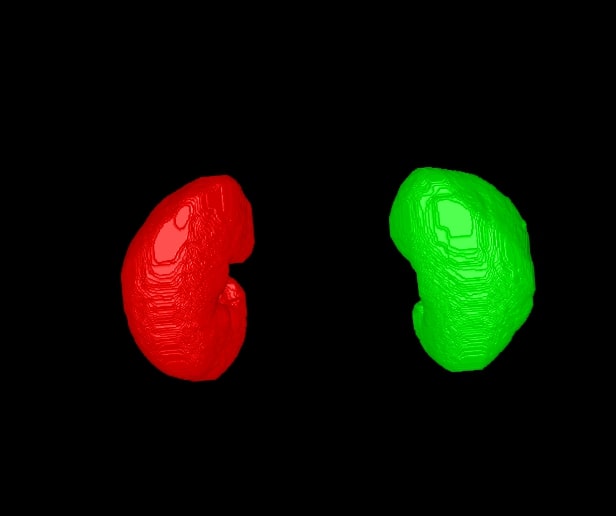}}
    \subfloat[U-Net]{\figB{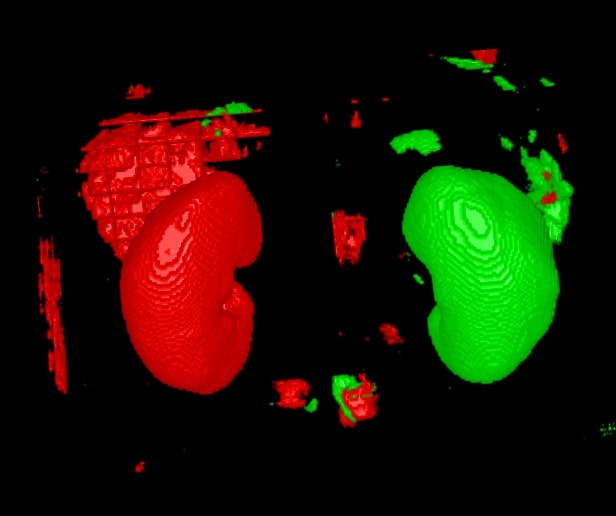}}\\[-1em]
    \subfloat[V-Net]{\figB{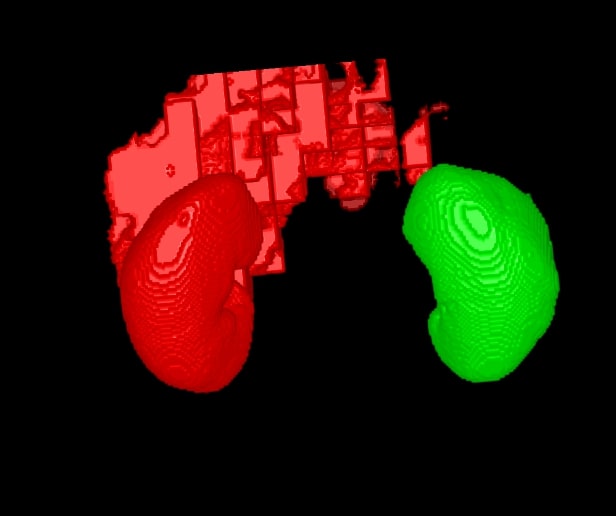}}
    \subfloat[Our proposed]{\figB{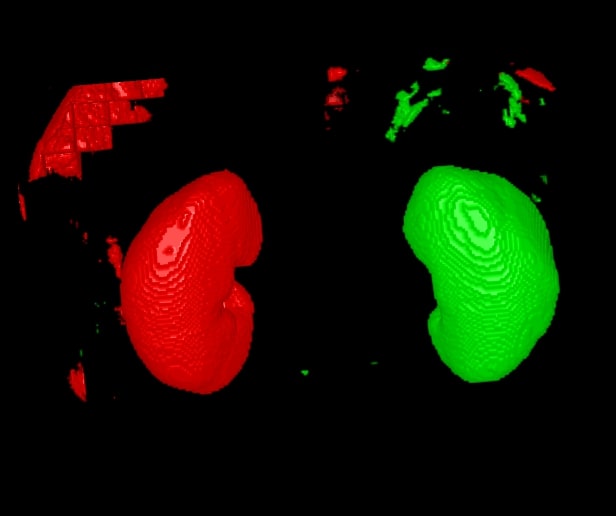}}
    \end{minipage}
    \caption{Two comparison examples. (a), (b), (c), and (d) denote ground truth and segmentation results of U-Net \cite{roth2018towards}, V-net \cite{milletari2016v}, our proposed network.}
    \label{fig:kidnet_seg_comp}
\end{figure}

The FCN model was first pre-trained on a multi-organ dataset of seven organ labels, including the liver, the spleen, the stomach, and the pancreas. This multi-organ dataset did not contain any kidney labels. Its latest segmentation $DSC$ is given in previous work \cite{shen2018influence}, which achieved an average accuracy of $87.3\%$ (excluding the background regions). The pre-training curve of our FCN model is plotted in Fig. \ref{fig:learningcurve}(a). The maximum {validation} $DSC$ score was $88.8\%$. The learning curve of the first fold's fine-tuning on the kidney dataset is shown in Fig. \ref{fig:learningcurve}(b). The maximum {validation} $DSC$ score of this fold was {$96.7\%$}. {To demonstrate the improvement made by introducing the spatial unit, we also evaluated the baseline U-Net \cite{roth2018towards}, i.e. the proposed architecture without the spatial unit. Furthermore, we implemented a variant U-Net architecture, V-Net \cite{milletari2016v}, for comparison on our kidney dataset. All comparison experiments have the same hyperparameter settings.} Validation results of these three methods are shown in Fig. \ref{fig:validation_curve}. The quantitative results of \added{unseen test data} of all 8-fold cross validations are shown in Table. \ref{tab:kidney_cv}. For reference, we also list several similar works on the kidney segmentation task, even though we used a different dataset and annotations. Two detailed segmentation examples using the proposed approach are shown in Fig. \ref{fig:kidney_seg}. Comparison segmentation results are shown in Fig. \ref{fig:kidnet_seg_comp}.

\newcommand{\splitcell}[3][c]{\begin{tabular}[#1]{@{}c@{}}#2\\[-0.2em]#3\end{tabular}}
\begin{table}[t]
\caption{Comparison of kidney segmentation methods. Three measurements, Dice coef., sensitivity, and Hausdorff distance, are shown with median[1st - 3rd quartile] or  mean $\pm$ standard values. CE-CT denotes contrast-enhanced CT, LOO denotes leave-one-out cross-validation, and 8-fold denotes 8-fold cross validation. All listed methods used own in-house dataset.}
\resizebox{\linewidth}{!}{%
\begin{tabular}{lllccccc}
\hline
\multirow{2}{*}{Method}  & \multirow{2}{*}{Modality} & \multirow{2}{*}{Case num}    & \multicolumn{2}{c}{$DSC$ (\%)}  & \multirow{2}{*}{$Se$ (\%)}   & \multicolumn{2}{c}{$HD$ (mm)}  \\
                                              &                 &                       & Left                      & Right           &              &  Left            &  Right            \\ \hline
\\[-1em]
\multicolumn{7}{c}{ (1) In-house dataset} \\ \hline\hline
\rowcolor{Gray}
Atlas-based random forest \cite{cuingnet2012automatic}  & CT   & \begin{tabular}[c]{@{}l@{}}Train: 233\\ Test: 179\end{tabular} &\splitcell{96.0}{[93.0-97.0]}  &\splitcell{96.0}{[93.0-97.0]}      & -        &$7.0\pm 10.0$  &$7.0\pm6.0$ \\
Atlas-based graph-cut \cite{chu2013multi}  & \begin{tabular}[c]{@{}l@{}}CT\\(portal-phase)\end{tabular}   &  100 (LOO)   & \multicolumn{2}{c}{$90.0\pm 5.0$}    & -   & \multicolumn{2}{c}{-}    \\
\rowcolor{Gray}
CNN+MSL \cite{zheng2017deep}  & \begin{tabular}[c]{@{}l@{}}CE-CT\\(multi-phase)\end{tabular} & \begin{tabular}[c]{@{}l@{}}Train: 370\\ Test: 78\end{tabular}  & \multicolumn{2}{c}{90.5}     & -        &\multicolumn{2}{c}{-}      \\
Shape-constrained Level-set \cite{skalski2017kidney} & CE-CT    & 10    & \multicolumn{2}{c}{86.2}     & -        &\multicolumn{2}{c}{19.6}   \\
\rowcolor{Gray}
2D patch-based CNN \cite{thong2016convolutional} & CE-CT   & \begin{tabular}[c]{@{}l@{}}Train: 79 \\ Test: 20\end{tabular}  &\splitcell{93.6}{[92.0-95.0]} &\splitcell{92.5}{[88.8-94.5]}   & \splitcell{93.78}{[90.7–95.6]}       & \splitcell{4.8}{[2.7-9.5]} &\splitcell{7.0}{[4.1-17.5]}   \\ \hline
\\[-1em]
\multicolumn{7}{c}{ (2) Our kidney dataset} \\ \hline\hline
\rowcolor{Gray}
Baseline U-Net \cite{roth2018towards}  &CE-CT &27 (8-fold)  &\splitcell{84.6}{[79.6-90.3]}  &\splitcell{91.0}{[84.5-92.7]} &\splitcell{91.1}{[83.1-94.2]} &\splitcell{12.0}{[7.3-15.4]} &\splitcell{5.2}{[4.8-8.0]} \\
V-Net \cite{milletari2016v}    &CE-CT &27 (8-fold)  &\splitcell{84.1}{[74.3-95.1]}  &\splitcell{86.7}{[74.6-92.9]} &\splitcell{85.6}{[81.5-94.5]} & \splitcell{13.9}{[7.5-20.8]} &\splitcell{8.1}{[5.2-28.1]} \\
\rowcolor{Gray}
Ours  & CE-CT      & 27 (8-fold)     & \splitcell{87.3}{[83.0-90.8]} &\splitcell{94.7}{[86.4-95.5]} &\splitcell{90.5}{[86.3-95.5]} & \splitcell{7.1}{[5.0-9.8]}  & \splitcell{4.6}{[3.8-6.9]}     \\ \hline
\end{tabular}}
\label{tab:kidney_cv}
\end{table}

\subsection{Renal artery segmentation}
 
We performed renal artery segmentation on the VOIs of the kidney regions that were extracted using the bounding-boxes of the segmented kidneys. Tumors were segmented manually. Instead of using a Dice coefficient, as in previous work, we presented a centerline overlap ($CO$) coefficient to evaluate the blood vessel segmentation accuracy. Using a blood vessel centerline to measure the segmentation accuracy was previously proposed \cite{metz20083d}. The Dice coefficient is sensitive to volume variation. In tiny blood vessel segmentation problems, geometric topology error is more important than volume error. The $CO$ coef. evaluates the overlapping of centerlines extracted from the ground-truth and segmentation results. Fig.~\ref{fig:centerline} describes how to calculate the overlapping ratio.  As demonstrated in previous work, the $CO$ coef. of the segmentation results exceeded $80\%$.

\subsection{Estimation of dominant regions}

\begin{figure}[tb]
\centering
\includegraphics[height=4.5cm]{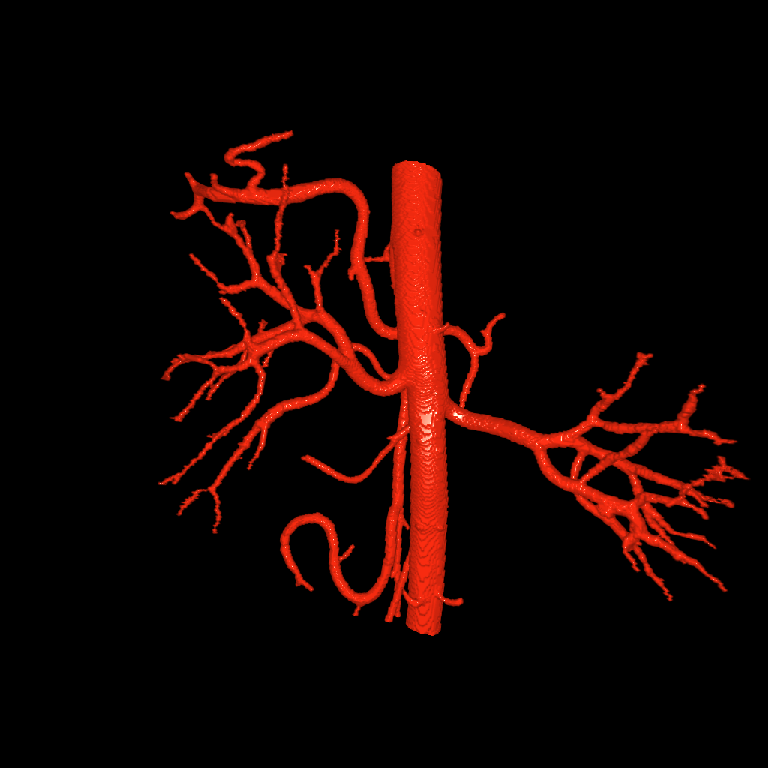}
\includegraphics[height=4.5cm]{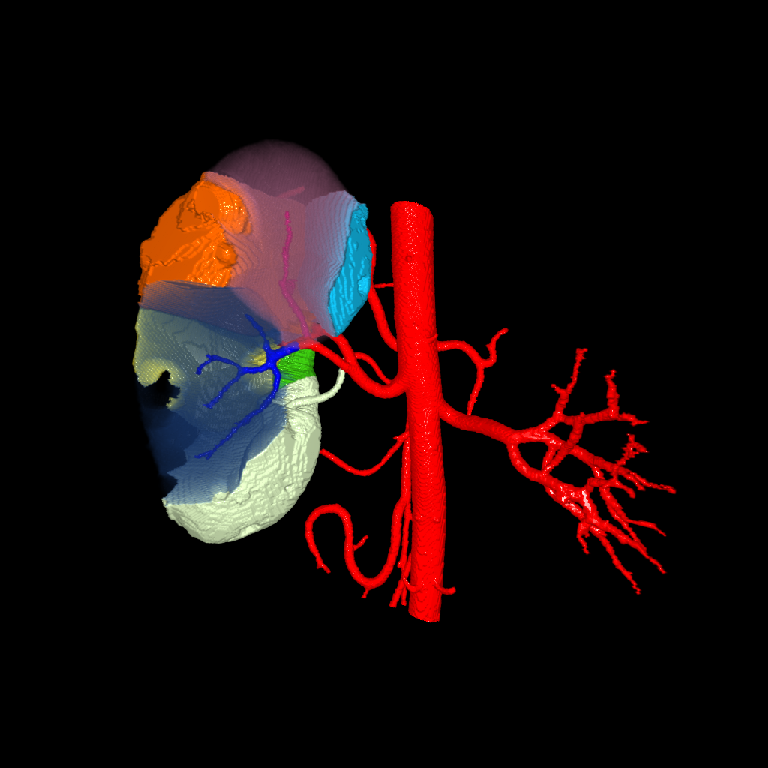}
\caption{Experimental result of Voronoi partition performed on original 3-D CT volume. Left: Abdominal blood vessels segmentation result. Thick blood vessels are segmented by region-grow semi-automatically, and renal arteries are extracted by proposed method. Right: Estimation result of renal vascular dominant regions that utilized Voronoi diagram.
}
\label{fig:voronoi}
\end{figure}

We conducted a quantitative evaluation of the estimation of dominant regions in eight cases involving 14 kidneys. We measured each estimated dominant region's Dice coef. with the ground truth. Since we cannot get the anatomical ground truth of the renal dominant regions, we used the ground truth of both kidney and renal artery to calculate a simulated ground truth of the dominant regions. The quantitative results are shown in Table \ref{tab:quant_ret}. One partition example on the original CT volume is shown in Fig.~\ref{fig:voronoi}. 

\begin{table}[]
\caption{Quantitative evaluation results of renal dominant regions. Eight cases were tested. Second and third rows present number of kidneys and dominant regions of each case. 4th row shows Dice score, and median values with 1st and 3rd quartiles values are demonstrated for each case.}
\resizebox{\columnwidth}{!}{%
\begin{tabular}{lccccccccc}
\hline\noalign{\smallskip}
 & Case 1 & Case 2 & Case 3 & Case 4 & Case 5 & Case 6 & Case 7 & Case 8 & Mean \\
\noalign{\smallskip}\hline\noalign{\smallskip}
Kidney \# & 2 & 2 & 2 & 2 & 1 & 1 & 2 & 2 &  \\
Regions \# & 12 & 22 & 16 & 10 & 12 & 7 & 10 & 14 &  \\
Dice coef. (\%) & \begin{tabular}[c]{@{}c@{}}81.3\\ {[}70.2 - 85.1{]}\end{tabular} & \begin{tabular}[c]{@{}c@{}}77.2\\ {[}66.9 - 87.8{]}\end{tabular} & \begin{tabular}[c]{@{}c@{}}79.7\\  {[}73.3 - 86.0{]}\end{tabular} & \begin{tabular}[c]{@{}c@{}}82.6\\ {[}71.9 - 87.0{]}\end{tabular} & \begin{tabular}[c]{@{}c@{}}82.2\\ {[}78.0 - 87.9{]}\end{tabular} & \begin{tabular}[c]{@{}c@{}}77.3\\ {[}73.0 - 81.2{]}\end{tabular} & \begin{tabular}[c]{@{}c@{}}80.0\\ {[}70.6 - 82.4{]}\end{tabular} & \begin{tabular}[c]{@{}c@{}}77.5\\ {[}70.4 - 87.2{]}\end{tabular} & \begin{tabular}[c]{@{}c@{}}79.9\\ {[}70.4 - 86.2{]}\end{tabular}\\
\noalign{\smallskip}\hline
\end{tabular}}
\label{tab:quant_ret}
\end{table}

\begin{figure}[tb]
\centering
\includegraphics[height=4.5cm]{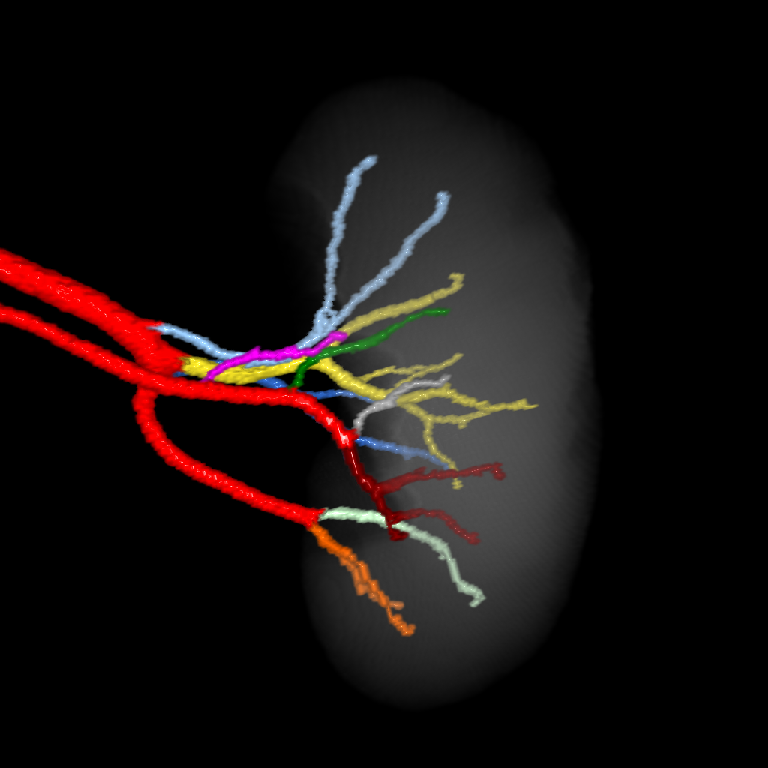}
\includegraphics[height=4.5cm]{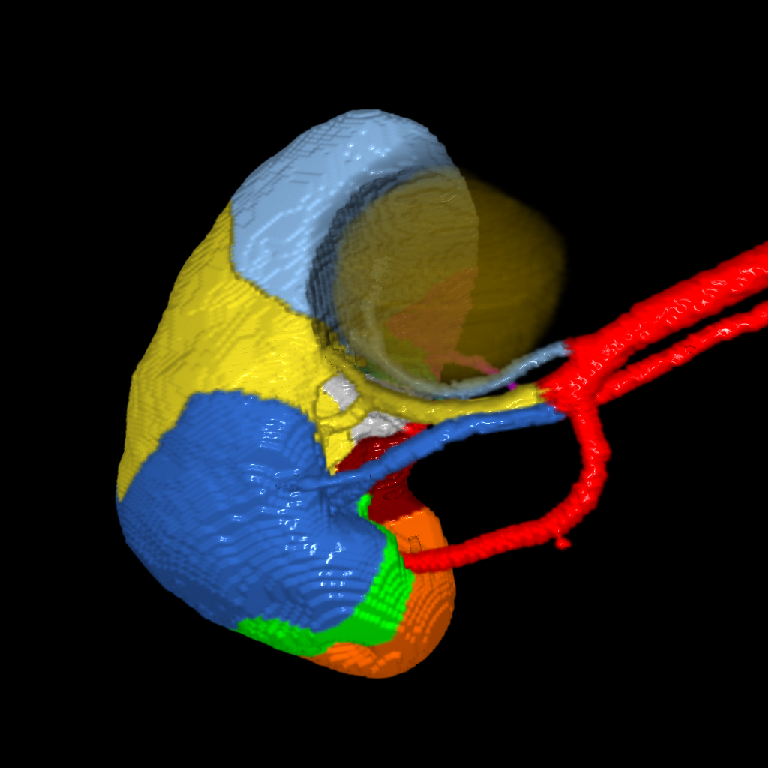}
\caption{Left: Segmented renal arteries labelled with different colors for each branch. Right: Estimation result of vascular dominant regions. Colors correspond to their dominated blood vessels. 5-mm margin was taken outside of tumor. Blue, yellow, green, and fuchsia regions are directly adjacent to a tumor.}
\label{fig:voronoi_validate}
\end{figure}
\section{Discussion}
In this work, we described a precise estimation approach for PN using a deep learning technique for kidney segmentation and a tensor-based graph-cut method for renal artery segmentation. We used our previously proposed ``Tensor-cut'' method for renal artery segmentation that can obtain over 80\% segmentation accuracy \cite{wang2016tensor}. For kidney segmentation, we presented an improved U-Net-like FCN architecture. The experimental results also demonstrate that better segmentation results were obtained by introduced the spatial information.

\subsection{Kidney segmentation}
As shown in Table \ref{tab:quant_ret}, compared with other U-Net-like architectures, our proposed spatially aware FCN achieved better segmentation results. Figure \ref{fig:kidnet_seg_comp} shows that the introduced spatial information effectively suppressed the FPs. Furthermore, our approach demonstrated competitive segmentation accuracy with related kidney segmentation methods. In this work, our $DSC$ and $Se$ are directly measured on the segmentation results without any post-processing. A 2D patch-based CNN method \cite{thong2016convolutional} performed post-processing, including opening, closing and extraction of two largest connected components. Therefore, a comparison of $HD$ is more useful for demonstrating our segmentation ability. Considering our limited dataset, we believe our proposed spatially aware FCN architecture \replaced{has potential}{is able} to achieve competitive results with state-of-the-art kidney-segmentation methods. Furthermore, the spatially aware unit can be easily incorporated in other architectures.

\begin{figure}[b]
\centering
\subfloat[]{\includegraphics[width=0.33\linewidth]{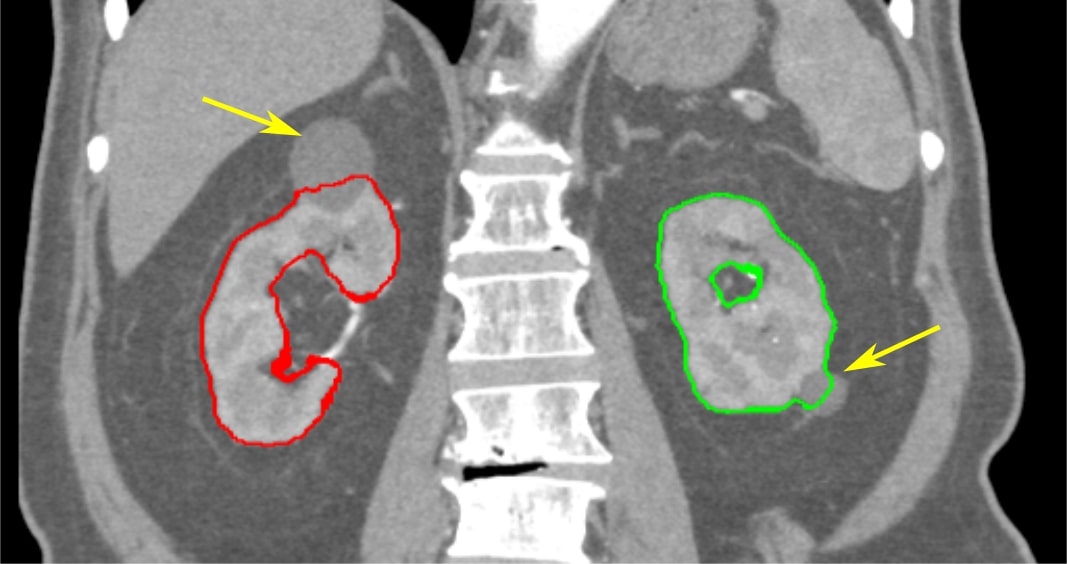}}
\subfloat[]{\includegraphics[width=0.33\linewidth]{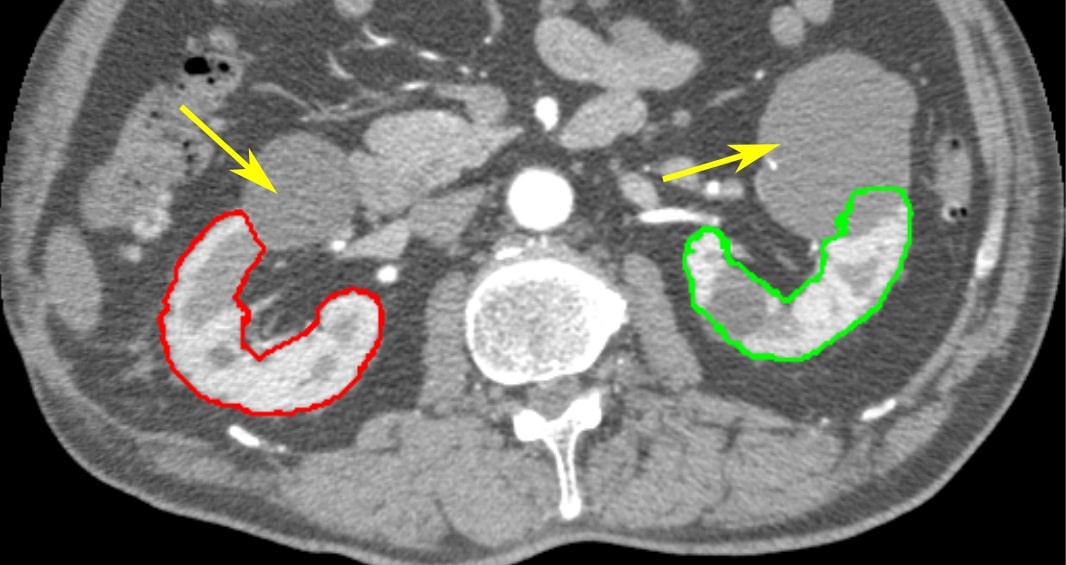}}
\subfloat[]{\includegraphics[width=0.33\linewidth]{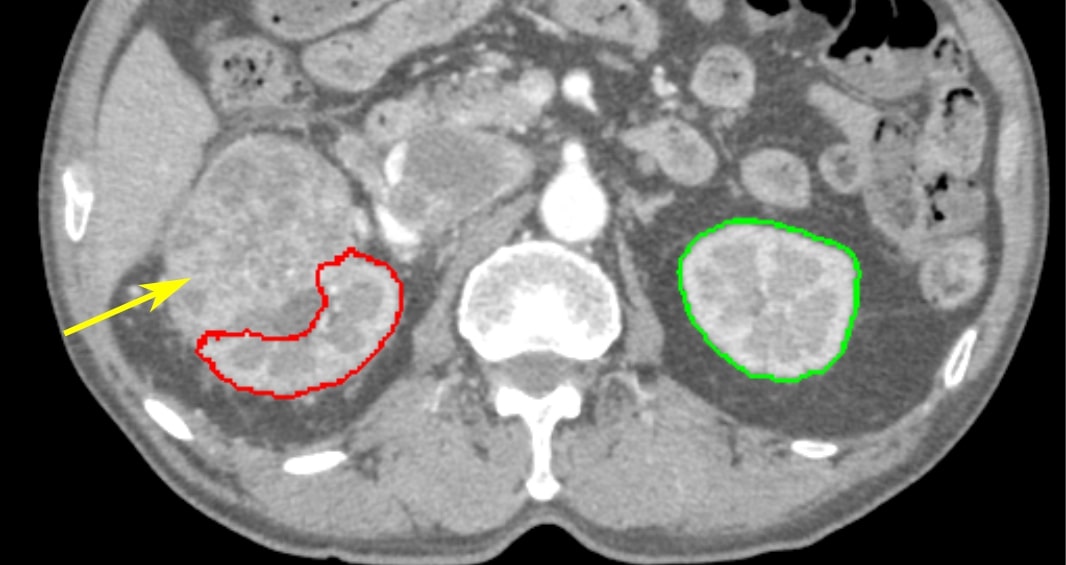}}
\caption{Limited segmentation performance: Yellow arrows indicate under-segmented regions. (a) and (b): limited segmentation performance on kidney cysts. (c): a case with kidney cancer at late stage.}
\label{fig:failure_ret}
\end{figure}

Although our network achieved good kidney segmentation results, its performance remains limited on such pathology patterns as kidney cysts and some late stage cancer. Several segmentation results are shown in Fig. \ref{fig:failure_ret}. Figs. \ref{fig:failure_ret}(a) and (b) show the under-segmented results of kidney cysts. Although the cyst regions have less effect on PN surgical plans, the segmentation of cysts can provide better diagnosis information. Fig. \ref{fig:failure_ret}(c) shows a case with late-stage cancer, and the proposed method failed to segment the whole kidney. One reason is that we have only one case that contains this pattern in our dataset; increasing the data with this pattern may improve the segmentation performance. {Among all \caseNum tested cases, two cases failed in segmentation ($DSC < 20\%$). This has also happened in other experiments using U-Net and V-Net. The major reason for this is that the slice thickness of these two failed cases is 2.0 mm, while that of all other cases ranges from 0.4 to 0.8 mm.}

\begin{figure}[tb]
\centering
\subfloat[]{\includegraphics[width=0.4\linewidth]{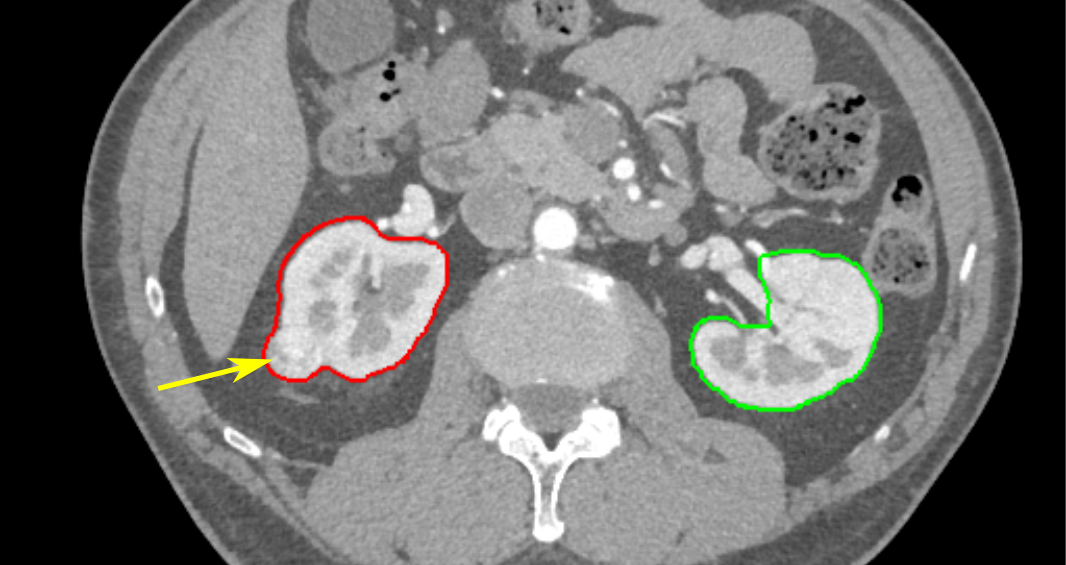}}
\subfloat[]{\includegraphics[width=0.4\linewidth]{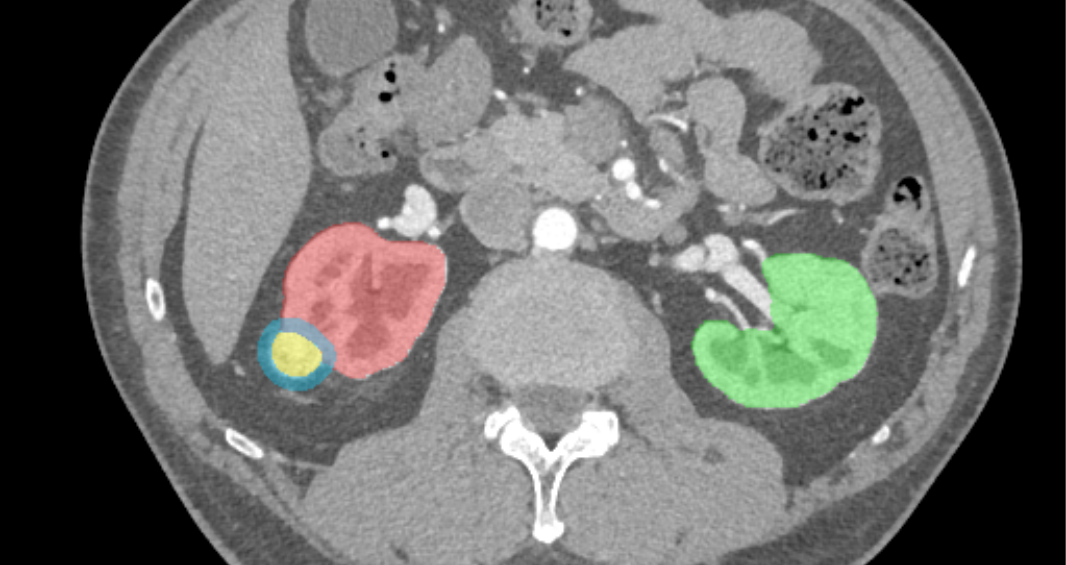}}
\caption{Manual tumor segmentation: yellow arrow indicates tumorous region. (a) automatic segmentation results of proposed methods: contour lines are shown in red and green for left and right kidneys. (b) manually segmented result of tumor: yellow region denotes tumor region with 5-mm margin surrounded shown in blue.}
\label{fig:tumor_seg}
\end{figure}

As shown in Fig. \ref{fig:tumor_seg}(a), the automatic segmentation of kidney regions contains the tumor region. Currently, we still need a manual segmentation process to extract tumors. One manually segmented result is shown in Fig. \ref{fig:tumor_seg}(b). A 5-mm margin was taken outside of the tumor for surgical safety. Much research has already been conducted on tumor segmentation using deep learning techniques \cite{kamnitsas2017efficient,havaei2017brain}. However, fine automatic segmentation of kidney tumors remains an essential future work.

\subsection{Renal artery segmentation}

\begin{figure}[hptb]
\centering
  \newcommand{\deffigureA}[1]{ \includegraphics[width=0.23\linewidth]{#1} }
  \newcommand{\deffigureB}[1]{ \includegraphics[width=0.3\linewidth]{#1} }
  \subfloat[2D axial slices of Case 6]{\deffigureB{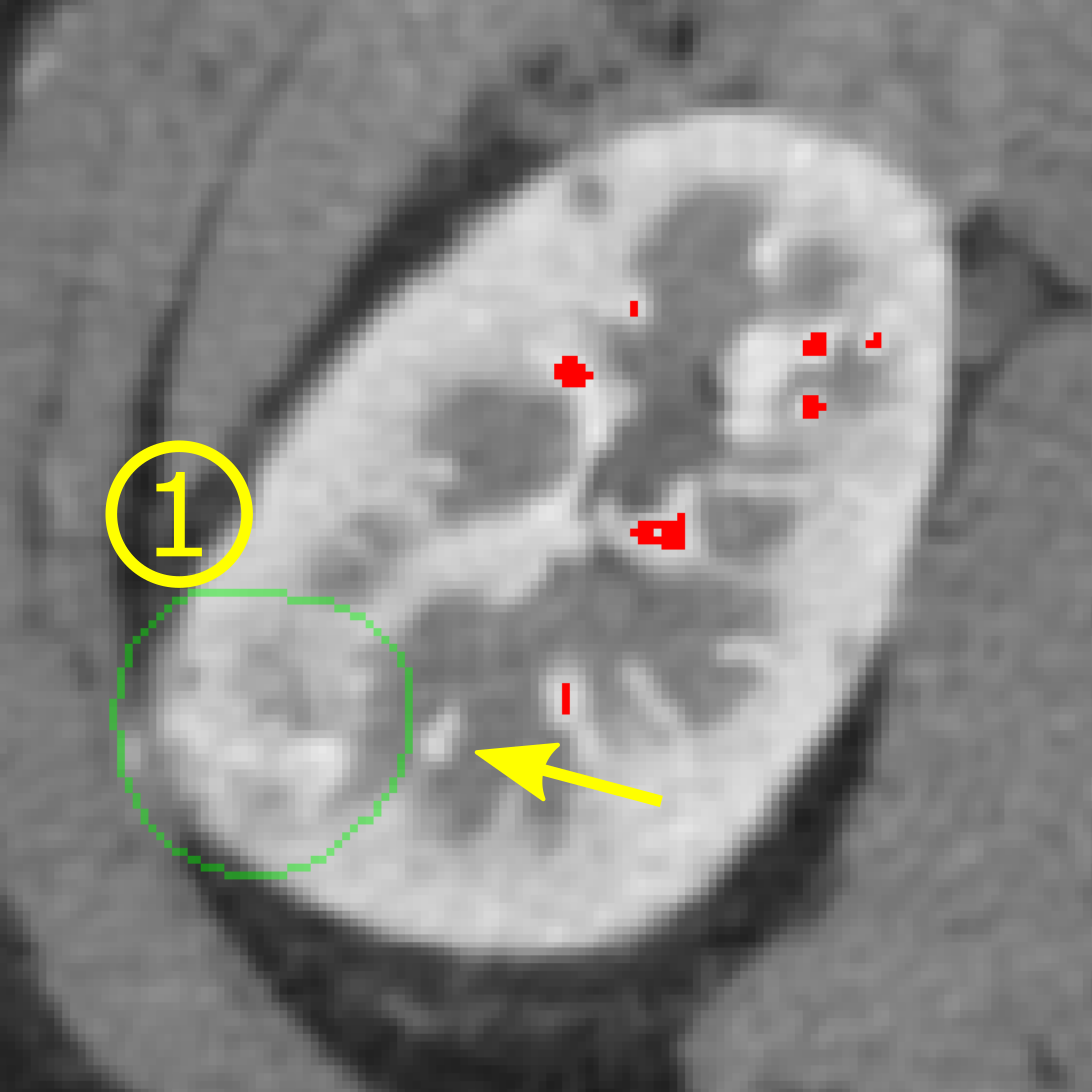}\deffigureB{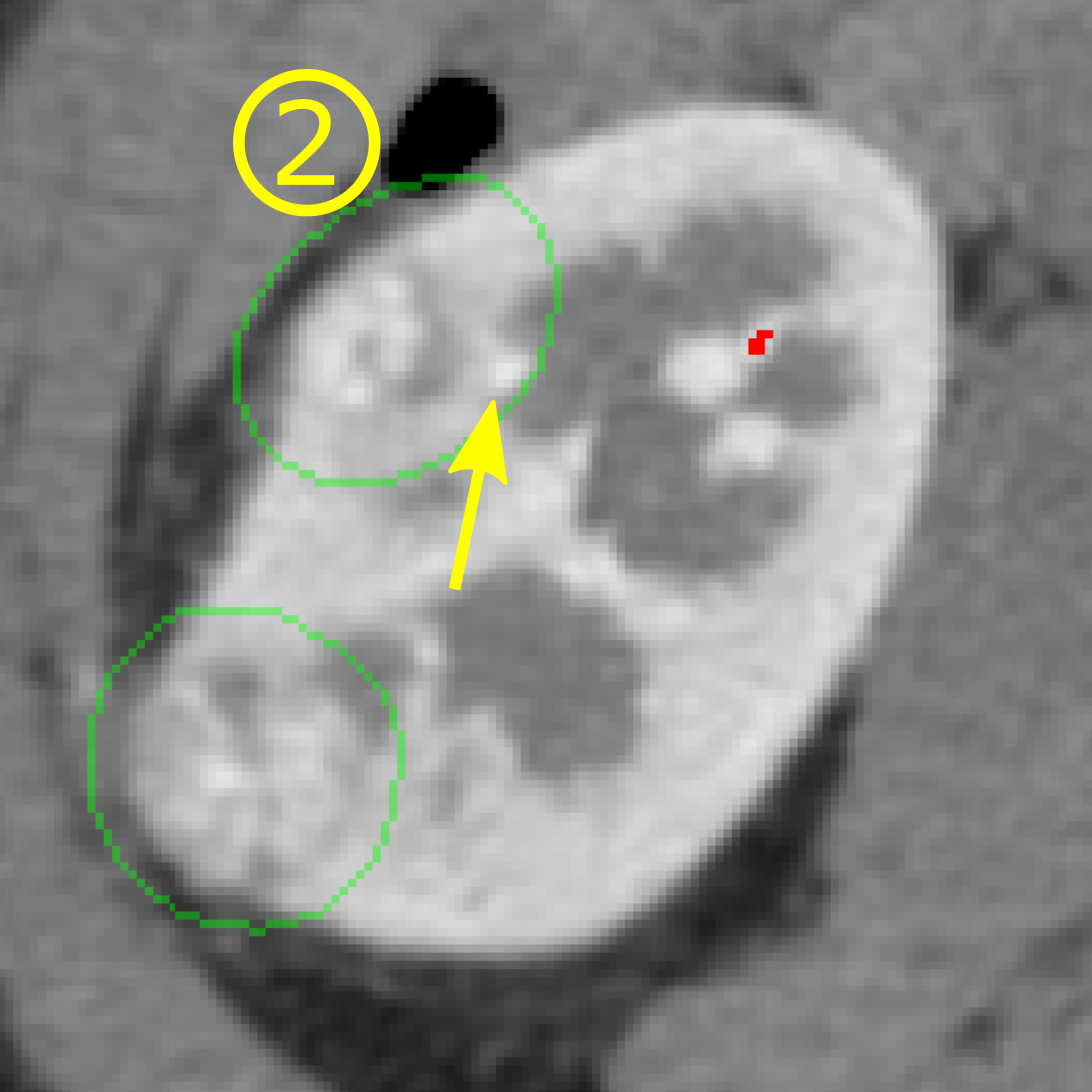}\deffigureB{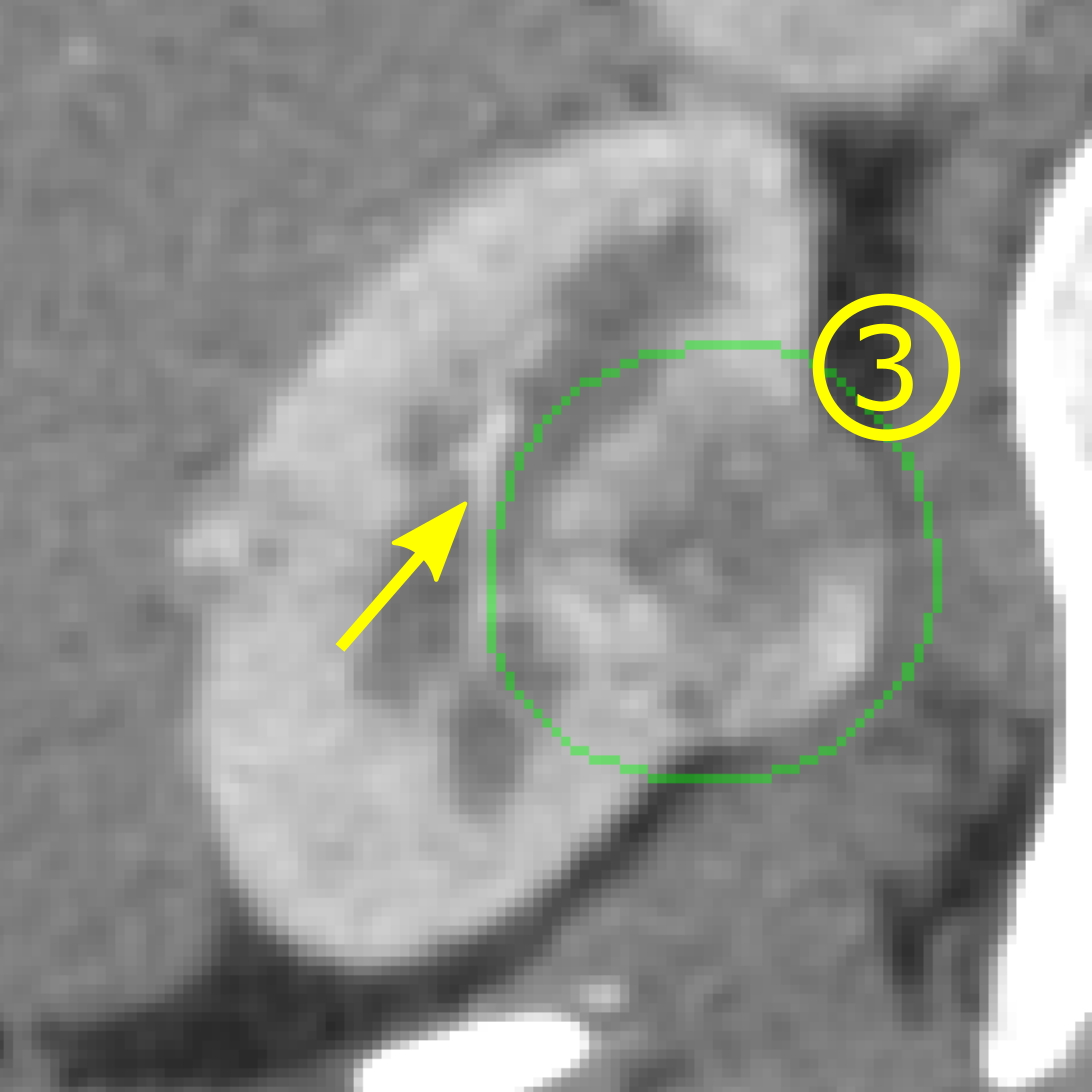}}\\
  \subfloat[Volume rendering of dominant regions of Case 6]{\deffigureA{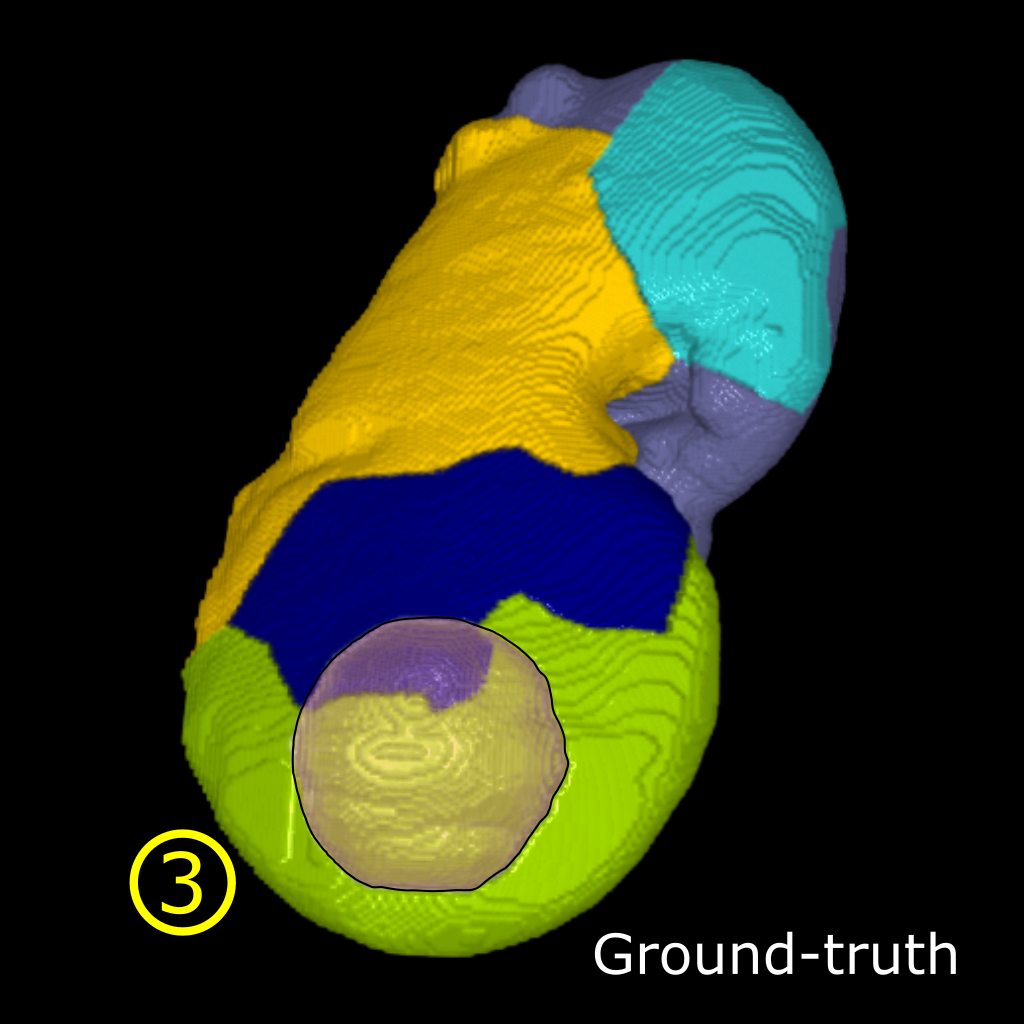}\hspace{-1em}\deffigureA{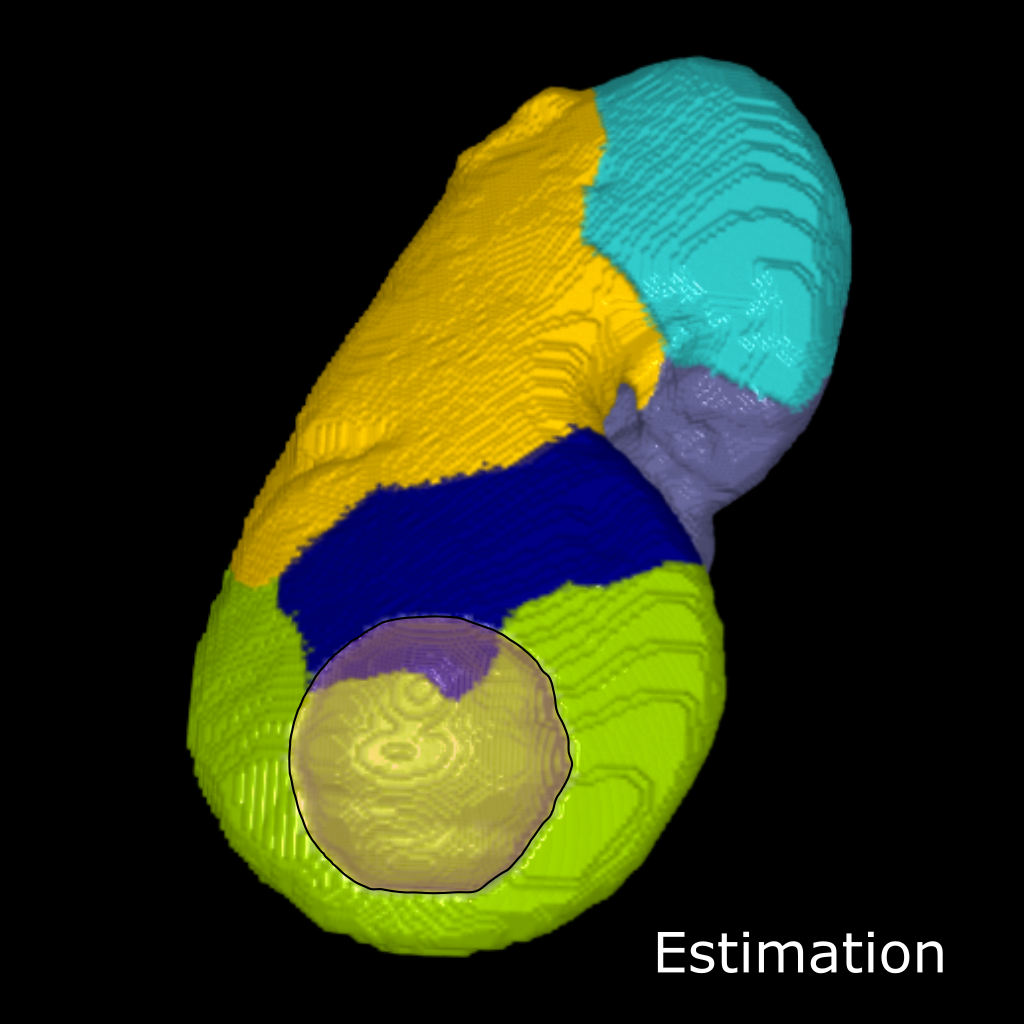}\deffigureA{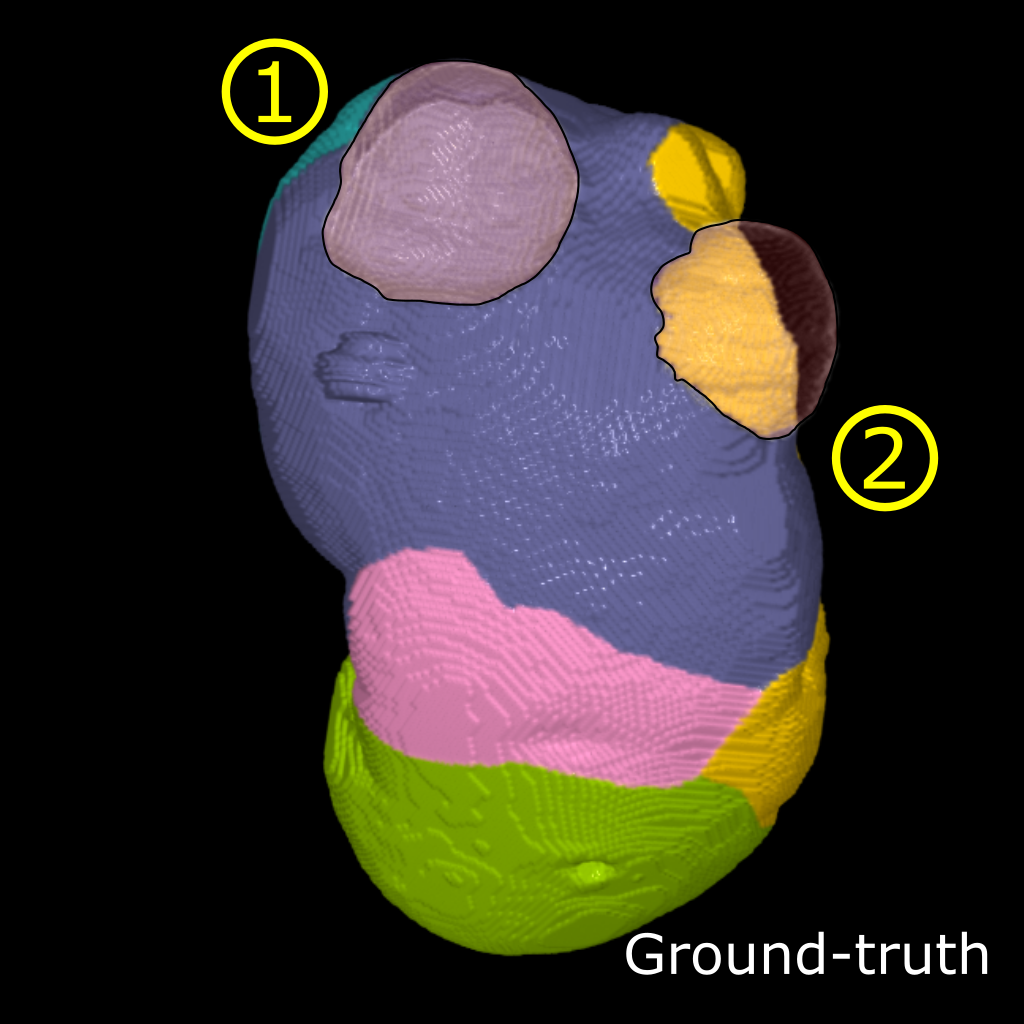}\hspace{-1em}\deffigureA{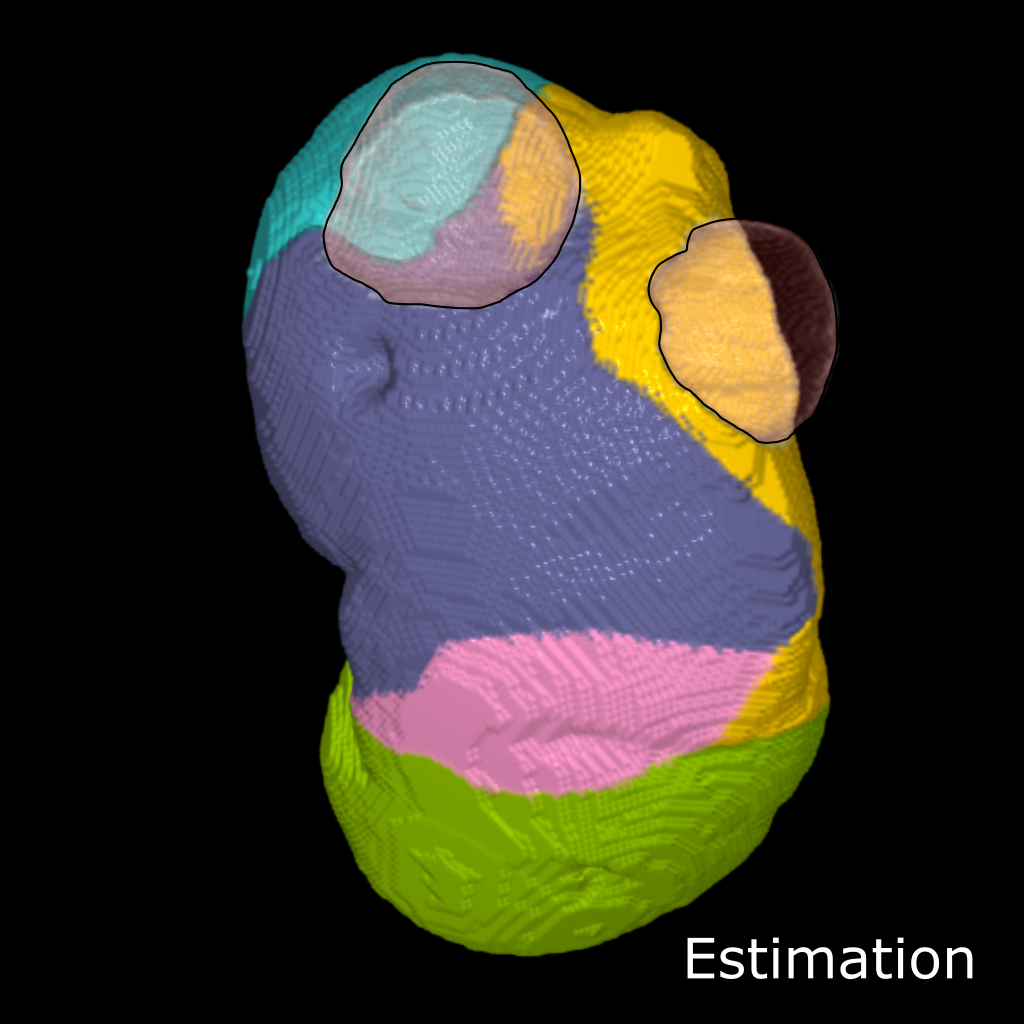}} \\
 \subfloat[2D sagittal slice of Case 2]{\deffigureB{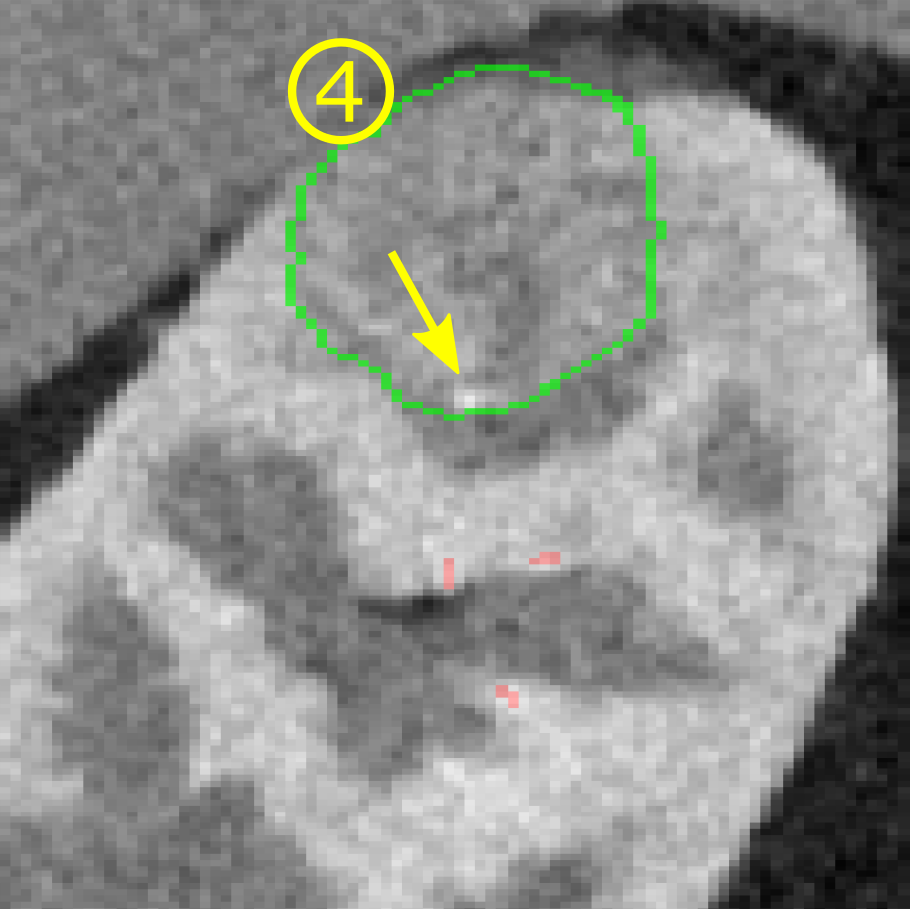}}
 \subfloat[Volume rendering of Case 2]{\deffigureB{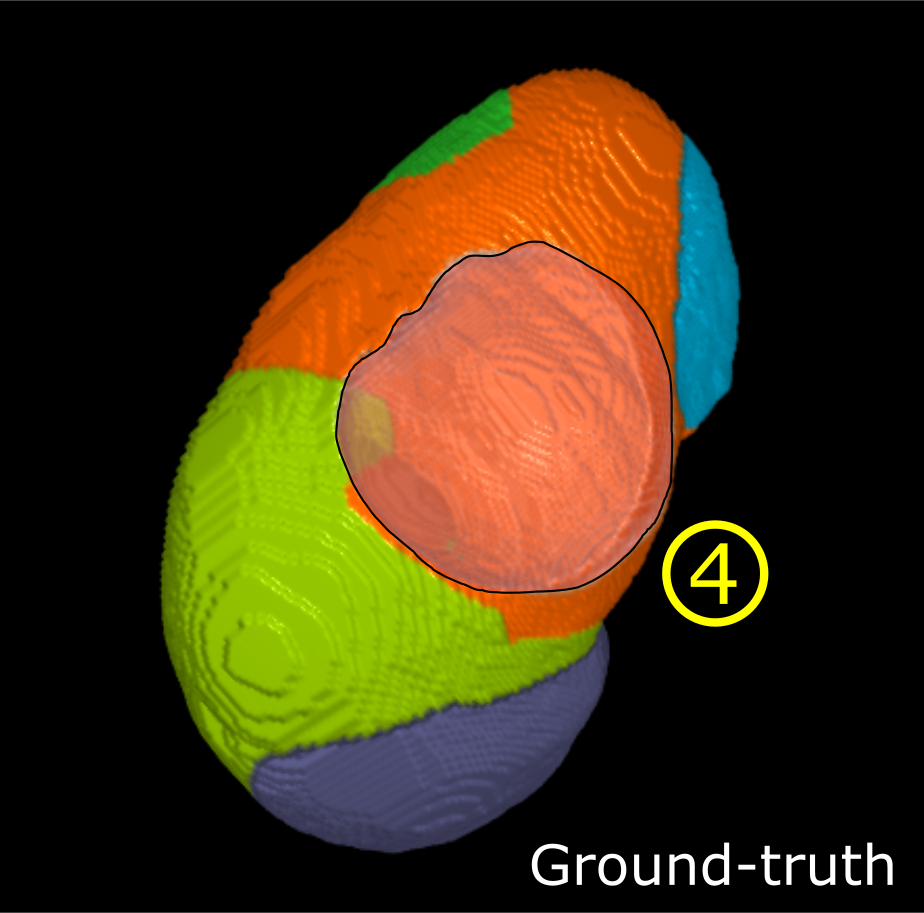}\hspace{-1em}\deffigureB{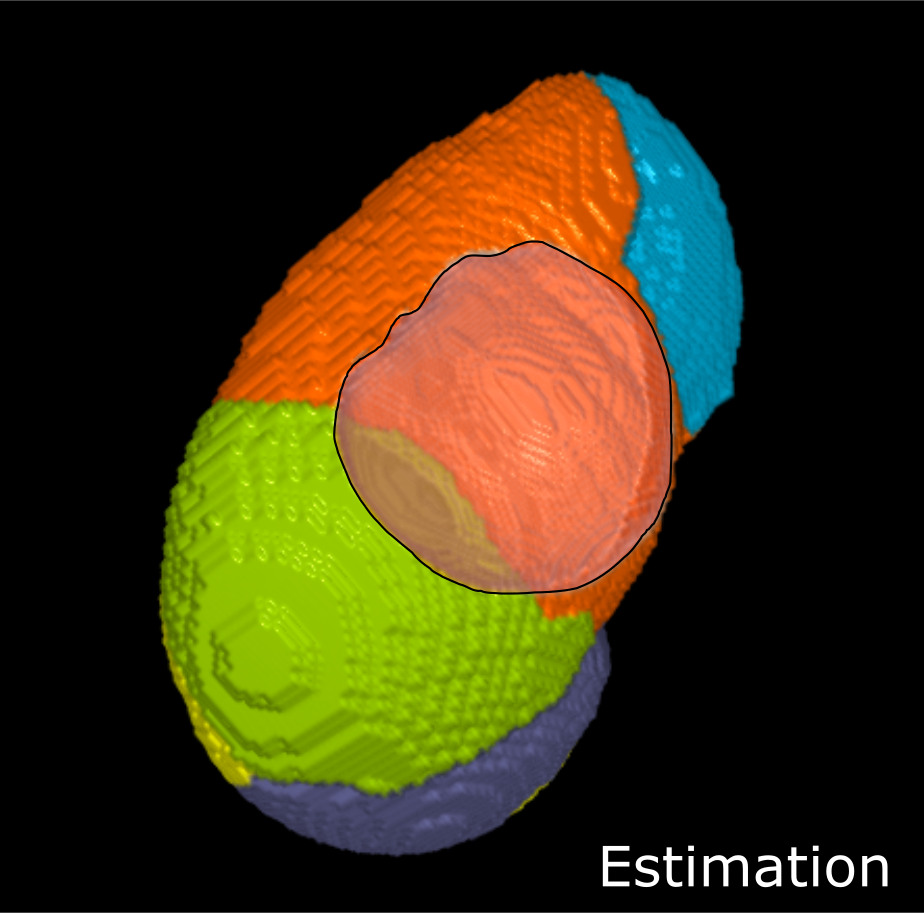}}
 \caption{Estimation results of dominant regions: Four tumors (1 in Case 2 and 3 in Case 6) are circled in green. Red regions in 2D slices indicate segmented renal arteries. Yellow arrows show under-segmented arteries around tumors. Volume rendering of each case is also shown in (b)(d). Contours of tumors are rendered in red with black contour line.}
 \label{fig:vessel_ref}
\end{figure}

As demonstrated in our previous work \cite{wang2016tensor}, the $CO$ coef. of renal artery segmentation exceeded 80\% and extracted about five generations of dichotomous branching that maximally extended from the abdominal aorta. This performance completely meets physician requirements for PN surgical planning. However, under-segmented renal arteries exist. As mentioned above, a renal artery is a critical step in our system. A variation of the segmented blood vessels will directly affect the result of the Voronoi diagram. We illustrated two examples with poor performance (Cases 2 and 6) in Fig. \ref{fig:vessel_ref} to show the influence. Yellow arrows indicate the under-segmented renal arteries around the tumors. In our experiments, the under-segmented renal arteries slightly affected the estimation results of the dominant regions around the tumors. For Case 2, the renal arteries dominating the orange and green regions are clamped the same as the ground-truth. However, for Case 6, the blue region is over-estimated. Although the segmentation accuracy still need to be improved, the renal artery segmentation performance for dominant-region estimation is acceptable in this work.

\begin{table}[tb]
\caption{Estimation result of renal vascular dominant regions: Nine regions were partitioned utilizing Voronoi diagram. Region colors refer to Fig.~\ref{fig:voronoi_validate}. Regions 1, 5, 6, and 7 are directly adjacent to tumors. \textit{Vol} denotes volume of dominant region, and \textit{Area} denotes adjacent area of dominant region and tumor. Ratio of both \textit{Vol} and \textit{Area} are given in next row.}
\resizebox{\columnwidth}{!}{%
\begin{tabular}{lglllggglll}
\hline\noalign{\smallskip}
      & 1 & 2 & 3 & 4 & 5 & 6 & 7 & 8 & 9 & all \\
      &Yellow &Blue &Light-green &Orange &Light-blue &Fuchsia &Green &Gray &Brown \\
\noalign{\smallskip}\hline\noalign{\smallskip}
$Vol$ ($mm^3$)  &49596.7   &22567.1   &22311.6   &14275.1   &20234   &1603.6   &8986.3   &460101   &16719.3   &160895    \\
$Vol$ Ratio (\%)      &\textbf{30.8}     &\textbf{14.0}     &\textbf{13.9}      &\textbf{8.9}       &\textbf{12.6}    &\textbf{1.0}      &\textbf{5.6}      &\textbf{2.9}      &\textbf{10.4}      &\textbf{100} \\
$Area$ ($mm^3$)  &17.6     &--        &--        &--   &761.5   &174.9   &22.5   &--   &--   &976.5    \\
$Area$ Ratio (\%) &\textbf{1.8}     &--        &--        &--   &\textbf{78.0}   &\textbf{17.9}   &\textbf{2.3}   &--   &--   &\textbf{100}    \\ 
\noalign{\smallskip}\hline
\end{tabular}
}
\label{tab:voronoi}
\end{table}{}

\subsection{Estimation of dominant regions}

\begin{figure}[tb]
\centering
\includegraphics[width=0.9\textwidth]{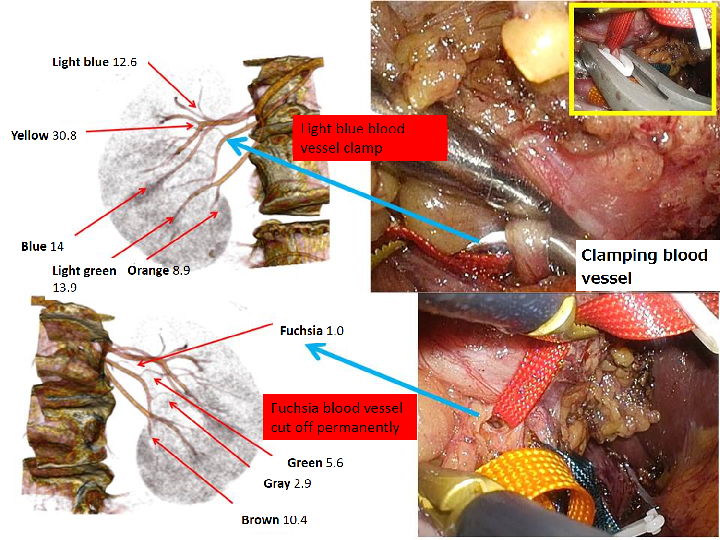}
\caption{Selective artery clamping is performed in nephrectromy surgery. Blood vessel in light blue shown in Fig.~\ref{fig:voronoi_validate} is clamped to prevent bleeding. Blood vessel in fuchsia is cut off permanently because it is the main blood vessel supply of nutrition to tumors.}
\label{fig:surgery}
\end{figure}{}

We utilized a Voronoi diagram to estimate the dominant regions of the renal arteries. {However, due to the limited size of annotated renal artery data, we could only perform quantitative validation on a small number of data (8 cases). The validation results from these 8 cases demonstrate that our estimation approach to the vascular dominant regions is generally correct. From this early study on the estimation of vascular dominant regions, we believe our proposed approach has the potential to improve estimation accuracy for dominant regions in clinical applications.} More quantitative clinical validations of the recovery from ischemic damage to normal kidney function can be found in our previous work \cite{yoshino2015}. 

%Surgical validation was performed on one case in this work. Two measures, area of adjacent surface and volume of region, were considered as main factors of blood vessel clamping scheme. PN surgical plan will designed using the information provided by our CAD system, such as a Table. \ref{tab:voronoi} and volume rendering figures Fig. \ref{fig:voronoi_validate}. More surgical validation remains a future work. It is also important to find the blood vessels which directly supply nutrition to tumor. Whether there is a way to find the blood vessel also become an interesting subject of future works.

% As a clinical application, one case in our validated dataset was used as a reference for preoperative surgical planning.

Figure~\ref{fig:surgery} shows the computerized analysis result for PN surgical planning. Since surgeons only compared the computerized analysis result with a PN surgery screen in one case, we show one comparison result in this figure. The segmented renal arteries of the left kidney and the corresponding dominant regions are shown in Fig.~\ref{fig:voronoi_validate}. 5-mm margin was taken outside of the tumor for surgical safety. Two measures were investigated: the volume of the vascular dominant region ($Vol$) and the area of the dominant region adjacent to tumor ($Area$). The quantitative result is shown in Table ~\ref{tab:voronoi}. Nephrectomy surgery was performed using a selective artery clamping scheme shown in Fig.~\ref{fig:surgery}. We confirmed that four regions are adjacent to the tumor, and thus at most four blood vessel branches should be clamped based on our simulated results. Surgeons clamped two blood vessel branches that dominate regions 5 and 6. An operation report shows that slight bleeding remains in regions 1 and 7. However, a valuable trade-off is found between surgical quality and residual renal function. From surgeon feedback, our proposed approach helped surgeons build preoperative surgical plans and focus on critical arteries during operations.

\section{Conclusion}
\replaced{This work presented a preliminary study on precise estimation approach for PN surgical planning.}{This work proposed a precise estimation approach for PN surgical planning.} Our approach mainly consists of three parts: a spatially aware fully connected convolutional network to extract kidney regions, a tensor-based graph-cut method to segment renal arteries, and a Voronoi diagram to estimate the dominant regions. The automatic kidney and renal artery segmentation methods achieved competitive results with state-of-the-art methods in our in-house data. The vascular dominant region is essential information for selective artery clamping, and its precise estimation will contribute to better ways of recovering residual renal function. \added{As a pilot study} on the estimation of renal vascular dominant regions, our experimental results in 8 cases demonstrated that our estimation approach achieved reasonable accuracy. However, more clinical evaluations \added{using large-scale database} are needed to prove the feasibility of our approach for clinical PN surgical planning.

%\begin{acknowledgements}
%If you'd like to thank anyone, place your comments here
%and remove the percent signs.
%\end{acknowledgements}

% BibTeX users please use one of
%\bibliographystyle{ieeetr}      % basic style, author-year citations
%\bibliographystyle{spmpsci}      % mathematics and physical sciences
%\bibliographystyle{spphys}       % APS-like style for physics
\bibliographystyle{elsarticle-harv}
\bibliography{ref}   % name your BibTeX data base

% \bibitem{pluto}
% ``http://pluto.s.m.is.nagoya-u.ac.jp", Nagoya University.

% \bibitem{yoshino}
% Y. Yoshino, T. Yamamoto, Y. Funahashi, M. Oda, C. Wang, M. Kagajo, K. Mori, M. Gotoh, ``Computational analysis of recovery from ischemic damage to kidney function undergoing robotic partial nephrectomy for renal tumor," Journal of Endourology, Vol. 29, Sup. 1, 2015

%\end{thebibliography}
\end{document}